\documentclass[sigconf]{acmart}

\renewcommand\footnotetextcopyrightpermission[1]{}
\acmConference[NBC 2026]{5th International Conference on Blockchain}{August 21--22, 2026}{Zurich, Switzerland}
\acmYear{2026}
\acmDOI{}
\acmISBN{}
\acmPrice{}

\hypersetup{
    colorlinks=true,
    linkcolor=blue!70!black,
    citecolor=blue!70!black,
    urlcolor=blue!70!black,
}

\usepackage{amsmath,amsfonts}
\usepackage{algorithm}
\usepackage{algpseudocode}
\floatname{algorithm}{Listing}

\usepackage{graphicx}
\usepackage{textcomp}
\usepackage{xcolor}
\usepackage{booktabs}
\usepackage{colortbl}
\usepackage{tabularx}
\usepackage{multirow}
\usepackage{placeins}   
\usepackage{needspace}  
\usepackage{fontawesome5}  

\newcommand{\AprThirtyMedOne}{2.25}
\newcommand{\AprThirtyStdOne}{1.09}
\newcommand{\AprThirtyMedTwo}{2.52}
\newcommand{\AprThirtyStdTwo}{0.92}
\newcommand{\AprThirtyDelta}{+12.3\%}
\newcommand{\NThirtyTwo}{10,763}
\newcommand{\EthThirtyTwo}{8.65M\,ETH}

\newcommand{\AprWindowMedOne}{2.61}
\newcommand{\AprWindowStdOne}{0.27}
\newcommand{\AprWindowMedTwo}{2.65}
\newcommand{\AprWindowStdTwo}{0.22}
\newcommand{\AprWindowDelta}{+1.5\%}
\newcommand{\NWindowTwo}{11,395}
\newcommand{\EthWindowTwo}{9.37M\,ETH}

\newcommand{\NAllOne}{898k}
\newcommand{\NAllTwo}{11,395}
\newcommand{\MedAllOne}{2.611}
\newcommand{\MedAllTwo}{2.651}
\newcommand{\DeltaAll}{+0.039}
\newcommand{\PAll}{1.8\!\times\!10^{-12}}
\newcommand{\NSoloOne}{180k}
\newcommand{\NSoloTwo}{7,932}
\newcommand{\MedSoloOne}{2.611}
\newcommand{\MedSoloTwo}{2.650}
\newcommand{\DeltaSolo}{+0.039}
\newcommand{\PSolo}{2.0\!\times\!10^{-3}}
\newcommand{\NProvOne}{719k}
\newcommand{\NProvTwo}{3,463}
\newcommand{\MedProvOne}{2.612}
\newcommand{\MedProvTwo}{2.652}
\newcommand{\DeltaProv}{+0.040}
\newcommand{\PProv}{8.8\!\times\!10^{-6}}

\newcommand{\DeltaAllPP}{+0.04}
\newcommand{\DeltaAllRel}{+1.5\%}

\newcommand{\NormSkew}{+0.50}
\newcommand{\NormW}{0.957}
\newcommand{\NormPval}{3.8\!\times\!10^{-36}}

\title{When Staking Rewards Compound: Measuring the impact of Ethereum's Pectra Upgrade}

\author{Mohammed Benseddik}
\affiliation{%
  \institution{University of Zurich}
  \city{Zurich}
  \country{Switzerland}
}
\email{mohammed.benseddik@uzh.ch}

\author{Benjamin Kraner}
\affiliation{%
  \institution{University of Zurich}
  \city{Zurich}
  \country{Switzerland}
}
\email{benjamin.kraner@uzh.ch}

\author{Claudio J.~Tessone}
\affiliation{%
  \institution{University of Zurich}
  \city{Zurich}
  \country{Switzerland}
}
\email{claudio.tessone@uzh.ch}

\begin{document}

\begin{abstract}
    Ethereum's beacon chain hosts over 920{,}000 active validators, a number inflated by the legacy 32\,ETH stake cap. The Pectra upgrade (May 2025) addresses this by introducing \texttt{0x02} compounding validators, raising the maximum stake per validator from 32 to 2{,}048\,ETH and enabling automatic reward reinvestment. This paper examines how compounding affects consensus-layer rewards, whether higher balances provide execution-layer advantages, and whether the APR uplift justifies migration for different staker types. We analyse adoption patterns across solo stakers and staking providers, investigate the role of consolidation (merging multiple 32\,ETH validators into one) in early migration, and identify barriers slowing the transition. Through simulation, we find that compounding provides roughly +5\% relative consensus-layer APR uplift for small balances, diminishing to under 1\% for large staking providers. Empirical analysis of all active beacon chain validators shows \texttt{0x02} validators achieving modestly higher median CL APR. Solo stakers show higher relative adoption but face operational barriers, whilst providers cite infrastructure costs and protocol constraints. The results suggest that without improved reward accessibility and stronger economic incentives, \texttt{0x02} migration will remain gradual despite its network efficiency benefits.
\end{abstract}


\maketitle

\noindent\textit{\small \href{https://github.com/benseddikmo/eth_pectra_paper_replication_package}{\faGithub~Replication package}~\cite{replication_package} (data, figures, and analysis code).}\smallskip

\section{Introduction}
\label{sec:intro}

Ethereum operates under a Proof-of-Stake (PoS) consensus mechanism that is widely regarded as more energy-efficient and accessible than Proof-of-Work (PoW). These advantages come with a more complex system of economic incentives designed to ensure correct and honest validator behaviour. In PoW systems such as Bitcoin, incentive-related concerns are largely limited to reward schedules \cite{eyal2018majority, kraner2022agent} or strategic attacks like block withholding \cite{schwarz2022stochastic}, with centralization emerging as the dominant long-term issue.

The transition to PoS introduced a fundamentally different incentive landscape. At the micro level, protocol parameters govern validator behaviour on a per-block basis, including mechanisms related to Maximal Extractable Value (MEV) and proposer-builder separation. At the macro level, the protocol shapes the network's monetary policy, staking rewards, and the structure of the staking ecosystem.

Empirical studies suggest that participation is broadly fair at the address level \cite{yan2025data}, though individual actors may still extract disproportionate value by exploiting MEV or protocol edge cases \cite{schwarz2022three, oz2023time, schwarz2023time}. At the macro level, more fundamental challenges arise, including sustainable inflation management, fairness across heterogeneous stakers, and effective risk management under fluctuating capital flows. The resulting techno-economic system is difficult to model \cite{jermann2023macro}, and ongoing innovation such as liquid staking and restaking further complicates its dynamics \cite{sugino2025analysis}.

Ethereum's beacon chain currently hosts over 920{,}000 active validators, a number inflated by the legacy 32\,ETH stake cap. Following the introduction of proof-of-stake, The Merge, and Shanghai-Capella withdrawals~\cite{ethereumForks2025}, the Prague-Electra (Pectra) upgrade (EIP-7251~\cite{eip7251}) addresses validator proliferation by raising the maximum effective balance to 2{,}048\,ETH through new \texttt{0x02} withdrawal credentials. This enables reward compounding and validator consolidation~\cite{eip7251,pectra_mainnet_announcement}.

For solo stakers, legacy \texttt{0x01} credentials sweep rewards above 32\,ETH to the withdrawal address rather than compounding, requiring over 11 years at current returns to accumulate enough stake for a new validator~\cite{ethresearch_maxeb}. With \texttt{0x02}, rewards compound directly in the beacon balance, reducing entry frictions~\cite{pectra_mainnet_announcement}. For large operators managing hundreds of thousands of validators, consolidation reduces key management overhead while preserving total stake~\cite{ethresearch_maxeb,pectra_scalability_impact}.

\begin{figure}[!t]
    \centering
    \includegraphics[width=\columnwidth]{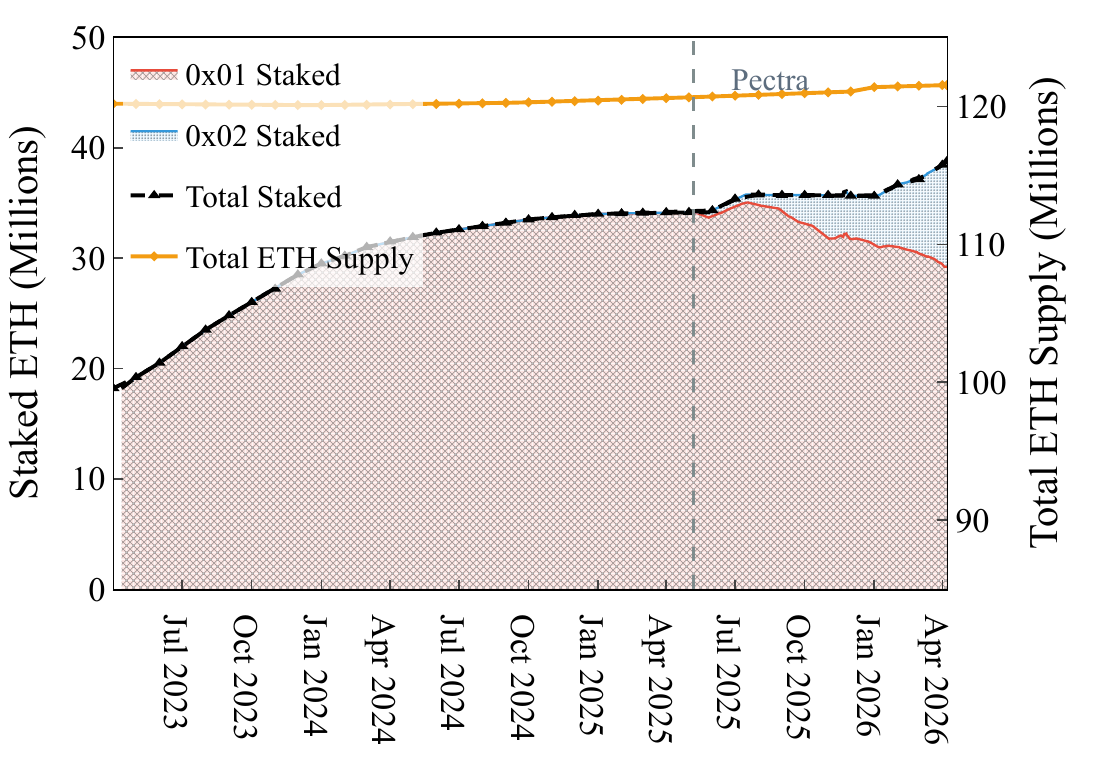}
    \caption{ETH staking since Shapella (Apr 2023). As of Apr 2026: 38.8M ETH staked (32\% of 121.6M supply); 9.7M ETH (25\%) on \texttt{0x02}. Source: Dune.}
    \label{fig:staked_eth_intro}
\end{figure}

Eleven months post-upgrade (Apr. 2026), adoption has reached 25\% of staked ETH (9.7M\,ETH, Fig.~\ref{fig:staked_eth_intro}), with new deposits dominating over consolidation. This raises three related research questions: (i) to what extent do \texttt{0x02} compounding validators improve staking returns in practice, and how does this effect vary by stake size; (ii) why has adoption remained limited despite the theoretical benefits; and (iii) which classes of stakers benefit most from \texttt{0x02}, and which face structural or operational disadvantages. We address these questions through a combination of simulations and empirical analysis.

The remainder of the paper is structured as follows: Sections~\ref{sec:overview}--\ref{sec:pos-changes} cover reward mechanics and simulations; Section~\ref{sec:empirical} presents the empirical analysis; and Sections~\ref{sec:threats}--\ref{sec:conclusion} discuss limitations and conclusions.

\subsection{Methodology \& Data}
\label{sec:methodology}

This study combines simulation-based analysis with empirical investigation. For the theoretical component, we employ Monte Carlo simulations to model validator balance evolution under various staking configurations, comparing \texttt{0x01} and \texttt{0x02} credential types across different stake sizes and time horizons. The empirical analysis draws on multiple data sources: Dune Analytics for on-chain staking metrics and validator entity classification, Beaconcha.in for consensus-layer performance data, and Etherscan for execution-layer reward attribution. Statistical comparisons of validator performance use annualised APR metrics with IQR-based outlier removal to ensure robustness.

\section{Overview of Ethereum Staking Rewards}
\label{sec:overview}

This section introduces the core notation used in the analysis. We decompose epoch-level validator rewards into consensus-layer (CL) and execution-layer (EL) components, specify the EIP-7251 effective-balance update rule, and summarise how effective balance determines proposer selection. These preliminaries form the baseline for the theoretical and empirical comparisons in subsequent sections. In particular, this section summarises the core mathematical expressions used throughout the analysis, defining \textbf{i.} total validator rewards, \textbf{ii.} effective balance updates under EIP-7251, and \textbf{iii.} proposer-selection probabilities.

\subsection{Balance Types}

Each validator maintains two balance quantities on the beacon chain. The \emph{actual balance} $B_{\text{act}}$ represents the validator's true holdings in Gwei, fluctuating continuously as rewards accumulate and penalties are incurred. The \emph{effective balance} $B_{\text{eff}}$ is a smoothed, discretised proxy used by the protocol to compute rewards, penalties, and proposer selection probabilities. This separation prevents minor balance fluctuations from causing constant recalculations of validator weights.

A validator becomes active and eligible for staking duties only when its actual balance reaches the \texttt{MIN\_ACTIVATION\_BALANCE} of \texttt{32\,ETH}, at which point the effective balance is initialised.

\begin{figure}[!t]
    \centering
    \includegraphics[width=\columnwidth,height=5.5cm]{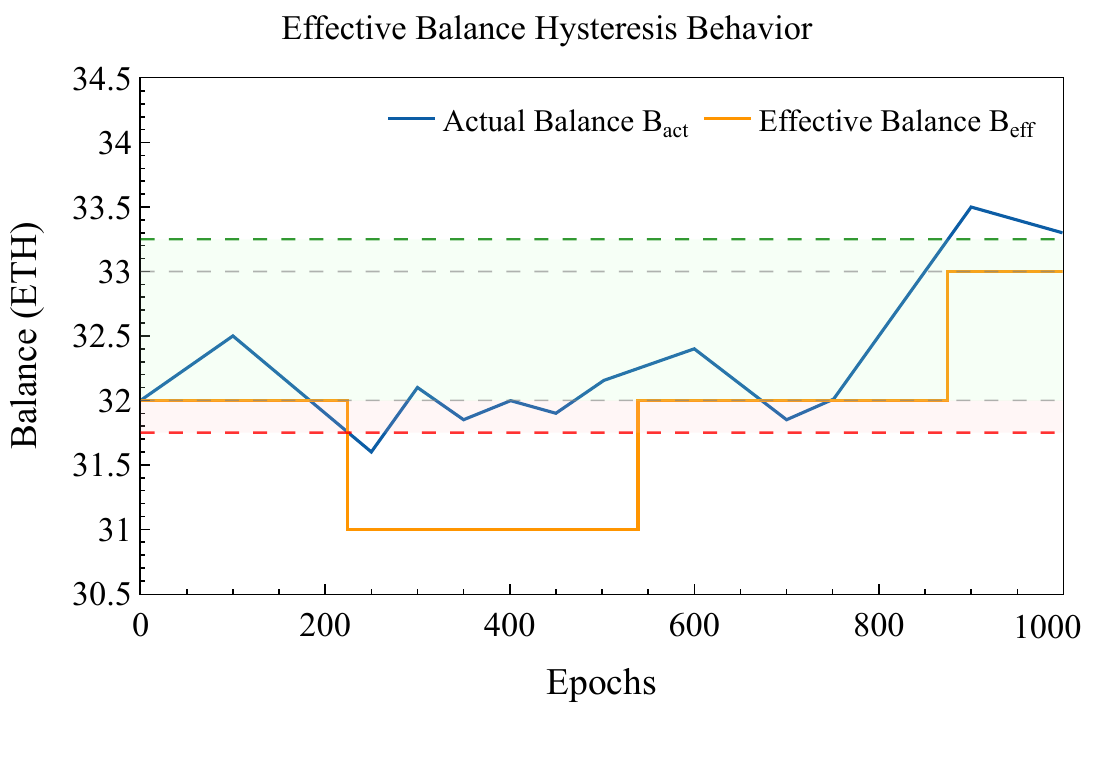}

    \caption{Hysteresis behaviour of effective balance $B_{\text{eff}}$ (orange) vs.\ actual balance $B_{\text{act}}$ (blue). The effective balance remains constant until $B_{\text{act}}$ crosses the downward threshold ($B_{\text{eff}} - \texttt{0.25\,ETH}$, red) or upward threshold ($B_{\text{eff}} + \texttt{1.25\,ETH}$, green), preventing frequent oscillations from small balance fluctuations.}
    \label{fig:effective-balance-hysteresis}
\end{figure}

The effective balance is bounded by a credential-dependent cap $B_{\text{cap}}$ (discussed later) and updated each epoch using a hysteresis mechanism\footnote{In Ethereum, \emph{hysteresis} refers to the protocol mechanism by which a validator's effective balance is updated only when its actual balance crosses predefined upward or downward thresholds, thereby preventing frequent oscillations in effective balance and stabilising reward calculations.}. The effective balance $B_{\text{eff}}$ is updated only when the actual balance $B_{\text{act}}$ exits the hysteresis band:
\begin{equation}
    B_{\text{eff}} \leftarrow
    \begin{cases}
        \min\bigl(\lfloor B_{\text{act}} \rfloor, B_{\text{cap}}\bigr), & \text{if } B_{\text{act}} \notin \bigl[B_{\text{eff}} - \Delta_{\downarrow},\, B_{\text{eff}} + \Delta_{\uparrow}\bigr] \\[4pt]
        B_{\text{eff}},                                                 & \text{otherwise}
    \end{cases}
\end{equation}
where $\Delta_{\downarrow} = \texttt{0.25\,ETH}$ and $\Delta_{\uparrow} = \texttt{1.25\,ETH}$ are the downward and upward hysteresis thresholds, respectively\footnote{EIP-8068 proposes revising these thresholds to address yield disparities~\cite{eip8068}.}. Fig~\ref{fig:effective-balance-hysteresis} illustrates the resulting staircase behaviour of $B_{\text{eff}}$ as $B_{\text{act}}$ grows.

\subsection{Reward Structure}
A validator's reward in a given epoch is the sum of CL and EL components:

\begin{equation}
    R_{v,\text{epoch}} = R_{\text{CL},v,\text{epoch}} + R_{\text{EL},v,\text{epoch}}
\end{equation}

Here, $R_{v,\text{epoch}}$ denotes the reward earned by validator $v$ in a given epoch, $R_{\text{CL},v,\text{epoch}}$ is the sum of all CL rewards and penalties accrued by $v$ during that epoch, and $R_{\text{EL},v,\text{epoch}}$ represents the total execution-layer rewards (priority fees and MEV) earned by $v$ in the same epoch.

Unless stated otherwise, any use of the term ``reward'' in the remainder of this paper refers to validator-level, epoch-level quantities of the form $R_{v,\text{epoch}}$ and its CL/EL components.

\subsubsection{Consensus-layer Rewards}
CL rewards $R_{\text{CL}}$ are the sum of all CL rewards and penalties in a given epoch:

\begin{equation}
    \begin{aligned}
        R_{\text{CL}}
         & =
        R_{\text{source}}
        + R_{\text{target}}
        + R_{\text{head}} \\
         & \quad
        + R_{\text{sync}}
        + R_{\text{proposer}}
        - \left(P_{\text{source}} + P_{\text{target}}\right)
    \end{aligned}
\end{equation}

Here, $R_{\text{source}}$, $R_{\text{target}}$, and $R_{\text{head}}$ denote attestation rewards for correct source, target, and head votes, respectively; $R_{\text{sync}}$ represents the sync committee reward; $R_{\text{proposer}}$ is the proposer reward for including attestations and sync committee messages; and $P_{\text{source}}$ and $P_{\text{target}}$ denote penalties incurred for missed or incorrect source and target votes.

Each \emph{CL} reward component earned by validator $v$ in a given epoch scales with a common base reward that depends on the validator's effective balance and the total active stake. Specifically, the reward for a given CL component can be written as
\begin{equation}
    R_{\text{component},v,\text{epoch}}
    =
    B_{\text{eff},v,\text{epoch}}
    \times
    \frac{F}{D \sqrt{S_{\text{total}}}}
    \times
    \frac{w_{\text{component}}}{64},
\end{equation}
where $B_{\text{eff},v,\text{epoch}}$ denotes the effective balance of validator $v$ in that epoch, $S_{\text{total}}$ is the total active stake on the network, and $F$ and $D$ are protocol constants defined in the Ethereum CL specifications~\cite{eth_consensus_specs,eth2book_base_reward}. The factor $w_{\text{component}}$ represents the fixed weighting associated with each CL reward type.\footnotemark

Since all other protocol parameters are validator-independent, a validator's total CL rewards in a given epoch satisfy $R_{\text{CL},v,\text{epoch}} \propto B_{\text{eff},v,\text{epoch}}$, i.e., they scale proportionally with the validator's effective balance.

\footnotetext{The weighting factors $w_{\text{component}}$ correspond to individual CL reward components (e.g., source, target, head, proposer, and sync committee rewards) and are fixed by protocol parameters.}

\subsubsection{Execution-layer Rewards}
Execution-layer rewards $R_{\text{EL}}$ are the sum of all execution-layer rewards (priority fees and MEV) in a given epoch. Unlike CL rewards, these are realised per slot where the validator is \emph{selected as block proposer} in the proof-of-stake protocol~\cite{ethereum_roadmap_pectra}; in other epochs, the expected execution income is zero.

\begin{equation}
    R_{\text{EL}} = R_{\text{priority\_fees}} + R_{\text{MEV}}
\end{equation}

Here, $R_{\text{priority\_fees}}$ denotes priority fees from included transactions and $R_{\text{MEV}}$ additional MEV revenue; execution-layer rewards are not analysed explicitly, as they are driven by proposer selection (i.e. they are proportional to the stake). We neglect any compounding associated with EL rewards.

\subsection{Performance Metrics}

The Annual Percentage Rate (APR) measures a validator's annualized yield based on the growth of its staking balance over a selected measurement window, using epoch-level granularity. APR is widely used as a standard metric in the industry to evaluate a validator's performance and capacity to generate rewards and return on investment (ROI).

The APR is defined as the period growth rate scaled to one year:

\begin{equation}
    \text{APR}_{v}
    =
    \left(
    \frac{R_{v,\text{window}}}{B_{\text{act,start},v}}
    \right)
    \times
    \left(
    \frac{E_{\text{year}}}{E_{\text{window}}}
    \right)
    \times 100
    \label{eq:apr-total}
\end{equation}

Here, $B_{\text{act,start},v}$ denotes the actual staking balance of validator $v$ at the beginning of the measurement window, $R_{v,\text{window}}$ is the total rewards accrued by validator $v$ during the window, $E_{\text{window}}$ is the number of epochs in the measurement window, and $E_{\text{year}} \approx 51{,}480$ denotes the number of epochs per year (with one epoch lasting approximately 6.4 minutes).

This quantity measures the annualized return on the validator's initial staking balance over the selected window. By restricting $R_{v,\text{window}}$ in~\eqref{eq:apr-total} to particular reward components (e.g., CL-only or EL-only rewards), we obtain the corresponding component-specific APRs.

APR denotes a \emph{simple} annual rate: it linearises the observed growth of a validator's balance over the measurement window to a one-year horizon using~\eqref{eq:apr-total}.\footnote{As a concrete example, if a validator starts a 90-day window with $B_{\text{act,start},v} = 100$~ETH and earns $R_{v,\text{window}} = 1$~ETH over that window (so $E_{\text{window}}/E_{\text{year}} \approx 1/4$), then $\text{APR}_{v} \approx (1/100)\times 4 \times 100 = 4\%$ (e.g., $\text{APR}_{\text{CL},v} = 3.5\%$ and $\text{APR}_{\text{EL},v} = 0.5\%$).}

\subsection{Withdrawal Credentials}
\label{sec:withdrawal-credentials}
Each validator has a 32-byte withdrawal credential; the first byte (\texttt{0x00}, \texttt{0x01}, or \texttt{0x02}) indicates the credential type. Validators are distinguished by this prefix: ``\texttt{0x01}'' validators pay rewards to an execution-layer address, while ``\texttt{0x02}'' validators pay rewards to an internal beacon-chain balance that can auto-compound. Both credential types existed before Pectra, but \texttt{0x02} validators only became relevant once EIP-7251 increased the maximum effective balance~\cite{eip7251}. Throughout this paper, we refer to ``\texttt{0x01} validators'' and ``\texttt{0x02} validators'' as shorthand for validators with the corresponding withdrawal credential type. \texttt{0x00} are legacy BLS credentials (Dec 2020) requiring conversion to \texttt{0x01} before withdrawals~\cite{eth_withdrawals_faq}; fewer than 1\% of validators remain on \texttt{0x00} as of April 2026. A future \texttt{0x03} type is proposed in EIP-7804~\cite{eip7804}. We exclude both \texttt{0x00} and \texttt{0x03} from this analysis.

\section{Proof-of-Stake (PoS) Related Changes Post-Pectra}
\label{sec:pos-changes}

The Pectra upgrade introduces a fundamental change to how validator rewards can grow over time. Building on the general framework for validator balances and rewards established earlier, this section examines how \texttt{0x02} validators enable automatic compounding of consensus-layer rewards, a mechanism previously unavailable to individual validators.

Post-Pectra, validators with \texttt{0x02} withdrawal credentials automatically \emph{compound} their CL rewards: income remains in the beacon-chain balance until the effective balance cap is reached. In contrast, \texttt{0x01} validators have their effective balance capped at \texttt{32\,ETH}. While the actual balance can temporarily exceed this cap (e.g., \texttt{32.23\,ETH}), the excess does not contribute to staking rewards since only the effective balance determines reward calculations. This surplus is periodically swept to the withdrawal address via automatic partial withdrawals, meaning \texttt{0x01} rewards do not compound on-chain.

We analyse the balance changes, compounding dynamics, the role of hysteresis in determining when effective balance increases, and the practical implications for validator profitability.

\subsection{Balance Changes}

EIP-7251 increased the \texttt{MAX\_EFFECTIVE\_BALANCE} while introducing a \texttt{MIN\_ACTIVATION\_BALANCE} of \texttt{32\,ETH} to preserve the lower bound for solo-stakers~\cite{eip7251}. The post-Pectra caps on effective balance are:

\begin{equation}
    B_{\text{cap}} =
    \begin{cases}
        \texttt{32\,ETH},   & \text{\texttt{0x01} validators} \\
        \texttt{2048\,ETH}, & \text{\texttt{0x02} validators}
    \end{cases}
    \label{eq:bcap}
\end{equation}

\subsection{\texttt{0x02} Compounding Mechanism}
\vspace*{-0.3\baselineskip}
The two reward layers differ not only in calculation but also in whether they compound (Table~\ref{tab:compounding}).

\begin{table}[b]
    \centering
    \caption{\texttt{0x02} validator rewards.}
    \label{tab:compounding}
    \begin{tabular}{lll}
        \toprule
        \textbf{Type}   & \textbf{Compounding}                           & \textbf{Destination}      \\
        \midrule
        $R_{\text{CL}}$ & Yes (cap: $B_{\text{cap}}=\texttt{2048\,ETH}$) & $B_{\text{act}}$ (beacon) \\
        $R_{\text{EL}}$ & No                                             & Withdrawal address        \\
        \bottomrule
    \end{tabular}
\end{table}

EL rewards are paid directly to the withdrawal address and do not compound. The impact of effective balance on block proposer probability is discussed later.

CL rewards drive compounding by increasing the validator's actual balance $B_{\text{act}}$. However, compounding only takes effect when $B_{\text{act}}$ crosses the hysteresis threshold and $B_{\text{eff}}$ increases by \texttt{1\,ETH}. Listing~\ref{box:compounding-mechanism} summarises this process.

\begin{algorithm}[htb]
    \caption{\texttt{0x02} CL Rewards Compounding Mechanism}
    \label{box:compounding-mechanism}
    \begin{algorithmic}[1]
        \While{$B_{\text{min}} \leq B_{\text{eff}} < B_{\text{cap}}$} $\Rightarrow$ Validator active, balance below cap
        \State CL rewards ($R_{\text{CL}}$) accumulate in $B_{\text{act}}$
        \State $B_{\text{act}}$ grows until it exceeds $B_{\text{eff}} + \texttt{1.25\,ETH}$
        \State $B_{\text{eff}} \gets B_{\text{eff}} + \texttt{1\,ETH}$ $\Rightarrow$ higher $R_{\text{CL}}$ ($R_{\text{CL}} \propto B_{\text{eff}}$)
        \EndWhile
        \State \textbf{Exit:} $B_{\text{eff}}$ hits $B_{\text{cap}}$ $\Rightarrow$ $R_{\text{CL}}$ still accrues in $B_{\text{act}}$ but no longer boosts future rewards
    \end{algorithmic}
\end{algorithm}
\noindent Here $B_{\text{min}} = 32$\,ETH for all validators; $B_{\text{cap}} = 32$\,ETH (\texttt{0x01}) or 2048\,ETH (\texttt{0x02}). When $B_{\text{eff}}$ reaches $B_{\text{cap}}$, CL rewards continue to accumulate in $B_{\text{act}}$, not the withdrawal address.

\subsection{Simulating the Compounding Advantage}
We compare a single \texttt{0x02} compounding validator against two \texttt{0x01} validators ($2 \times 32$\,ETH, rewards unstaked) over 730 days, starting at 64.5\,ETH total stake with 99\% participation rate. Swept \texttt{0x01} rewards are not restaked, as the validator would need to accumulate at least 32\,ETH to activate a new validator. We also do not model $B_{\text{eff}}$ gaming strategies; see EIP-8068~\cite{eip8068} for discussion of hysteresis optimisation.

The simulation models all CL reward components under fixed network conditions, with total active stake held constant at 38.8M\,ETH (as of April 2026). In practice, staked ETH fluctuates (${\sim}$34M to ${\sim}$38.8M\,ETH during 2025--2026, representing ${\sim}$14\% variation~\cite{beaconchain}), but this simplification does not materially affect relative comparisons. We exclude execution layer priority fees from block proposals to provide a conservative baseline; inclusion would improve absolute returns but depends heavily on proposal frequency, which is probabilistic and validator-count dependent.

\begin{figure}[!t]
    \centering
    \includegraphics[width=0.95\columnwidth]{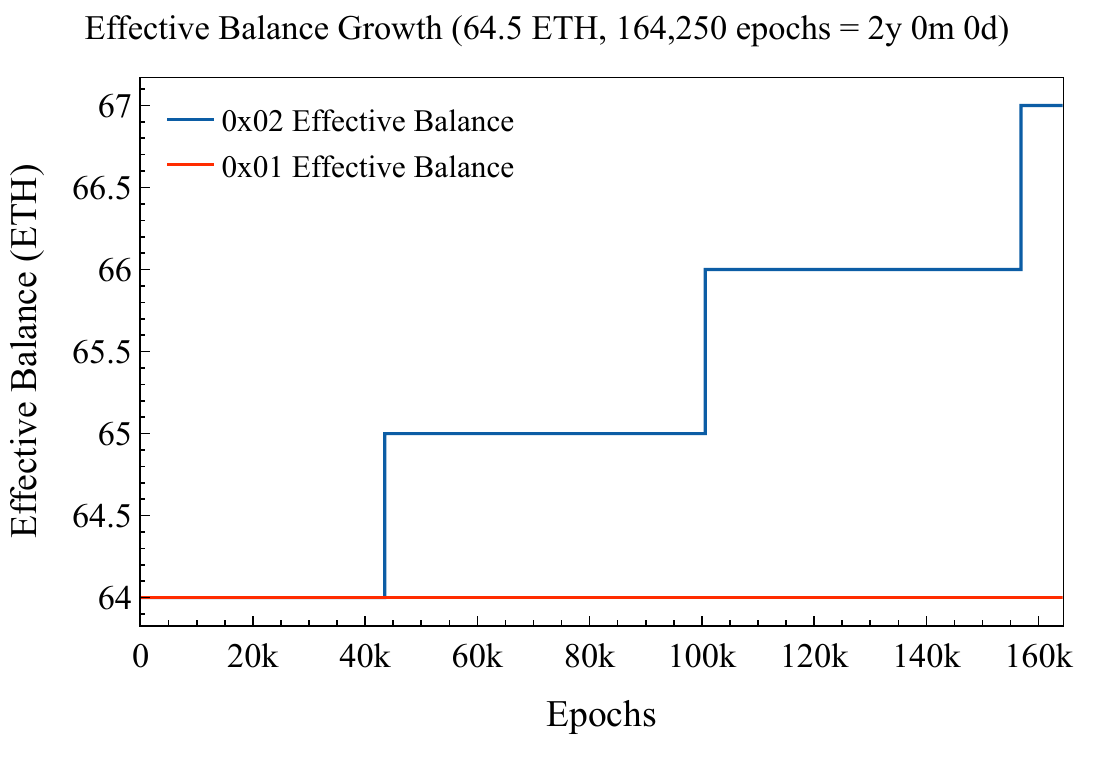}
    \vspace{-2mm}
    \caption{Effective balance evolution over 2 years: \texttt{0x02} grows via compounding whilst \texttt{0x01} validators remain at their 32\,ETH caps.}
    \label{fig:compounding-effective-balance}
\end{figure}

Fig~\ref{fig:compounding-effective-balance} illustrates the key difference: \texttt{0x02} effective balance grows from 64 to 67\,ETH through discrete hysteresis-driven increments, whilst \texttt{0x01} remains capped at 32\,ETH per validator. Each 1\,ETH increment in $B_{\text{eff}}$ increases future attestation rewards proportionally, creating a positive feedback loop.

\begin{figure}[!t]
    \centering
    \includegraphics[width=0.95\columnwidth]{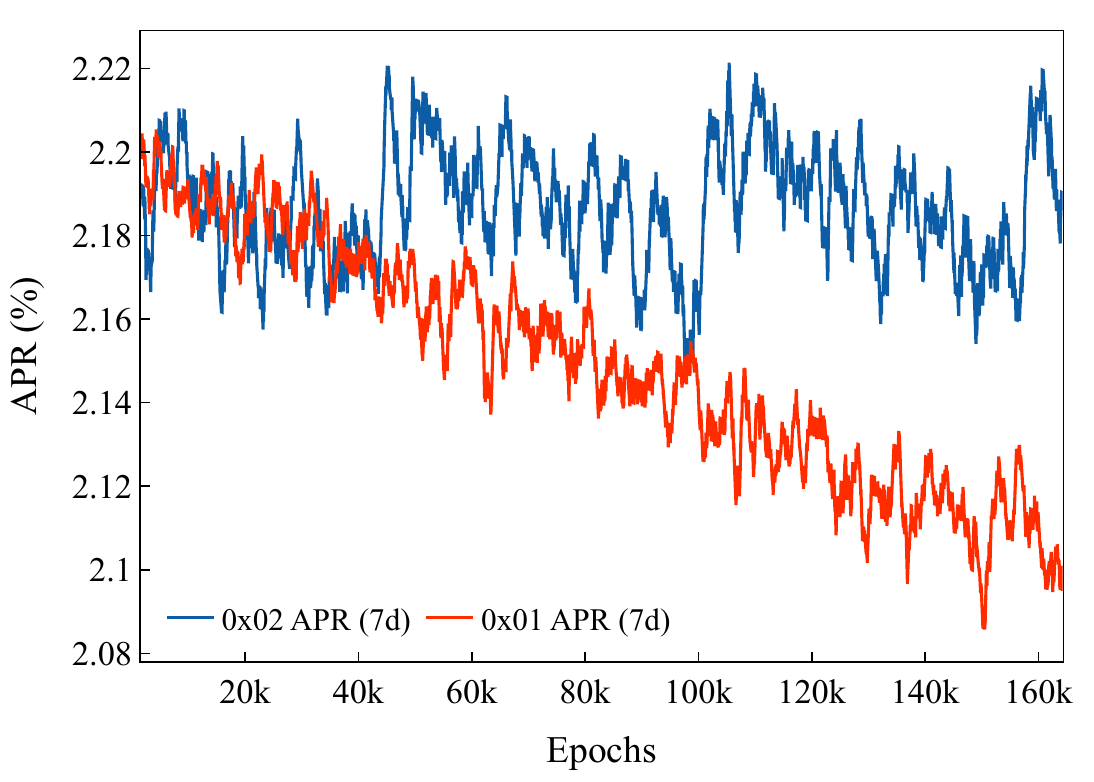}
    \vspace{-2mm}
    \caption{Rolling APR (top) and $\Delta$APR (bottom) over 164,250 epochs. The APR gap widens initially as \texttt{0x02} $B_{\text{eff}}$ grows, then stabilises around 2\,bps.}
    \label{fig:compounding-apr}
\end{figure}

Fig~\ref{fig:compounding-apr} quantifies the APR impact. We calculate APR using total portfolio ETH (staked balance plus swept rewards), so \texttt{0x01} APR decreases over time as idle ETH accumulates whilst waiting to reach 32\,ETH for a new validator. In Year~1, \texttt{0x02} achieves 2.21\% APR versus 2.20\% for \texttt{0x01} (${\sim}$2\,bps difference, ${\sim}$1.9\% relative CL reward uplift). By Year~2, the gap widens as $B_{\text{eff}}$ continues growing for \texttt{0x02}. Table~\ref{tab:compounding-results} summarises the full 2-year results.

The compounding advantage is bounded by two factors: (1) the slow rate of $B_{\text{eff}}$ growth due to hysteresis thresholds, and (2) the diminishing marginal impact of each additional ETH on percentage returns. The next section examines how this advantage scales across different stake sizes.
\begin{table}[h!]
    \centering
    \begin{tabular}{@{}lll@{}}
        \toprule
        \textbf{Metric} & \textbf{Value} & \textbf{Unit} \\
        \midrule
        Initial Balance & 64.5 & ETH \\
        Simulation Period & 730 & days \\
        Participation Rate & 99.0 & \% \\
        \addlinespace[0.5em]
        \textbf{0x02 Compounding} &  & \\
        \quad Validators (initial) & 1 & \\
        \quad Validators (final) & 1 & \\
        \quad Final Balance & 67.3848 & ETH \\
        \quad Final Effective Balance & 67 & ETH \\
        \quad Total Rewards & 2.8848 & ETH \\
        \quad Balance Growth & 4.47 & \% \\
        \quad APR (Year 1) & 2.2134 & \% \\
        \addlinespace[0.5em]
        \textbf{0x01 Non-Compounding} &  & \\
        \quad Validators (initial) & 2 & \\
        \quad Validators (final) & 2 & \\
        \quad Final Balance & 67.3314 & ETH \\
        \quad Final Effective Balance & 64 & ETH \\
        \quad Total Rewards & 2.8314 & ETH \\
        \quad Balance Growth & 4.39 & \% \\
        \quad APR (Year 1) & 2.1950 & \% \\
        \addlinespace[0.5em]
        \textbf{Difference (0x02 - 0x01)} &  & \\
        \quad Extra Rewards & 0.0534 & ETH \\
        \quad Reward Advantage & 1.89 & \% \\
        \quad APR Difference & 0.0184 & \% \\
        \bottomrule
    \end{tabular}
    \caption{Simulation results comparison: 0x02 compounding vs.\ 0x01 non-compounding validators (64.5\,ETH initial balance, 730 days).}
    \label{tab:compounding-results}
\end{table}

\subsection{Scaling Effects Across Balance Ranges}
\label{sec:scaling-effects}

\begin{figure}[!t]
    \includegraphics[width=\columnwidth,height=6cm]{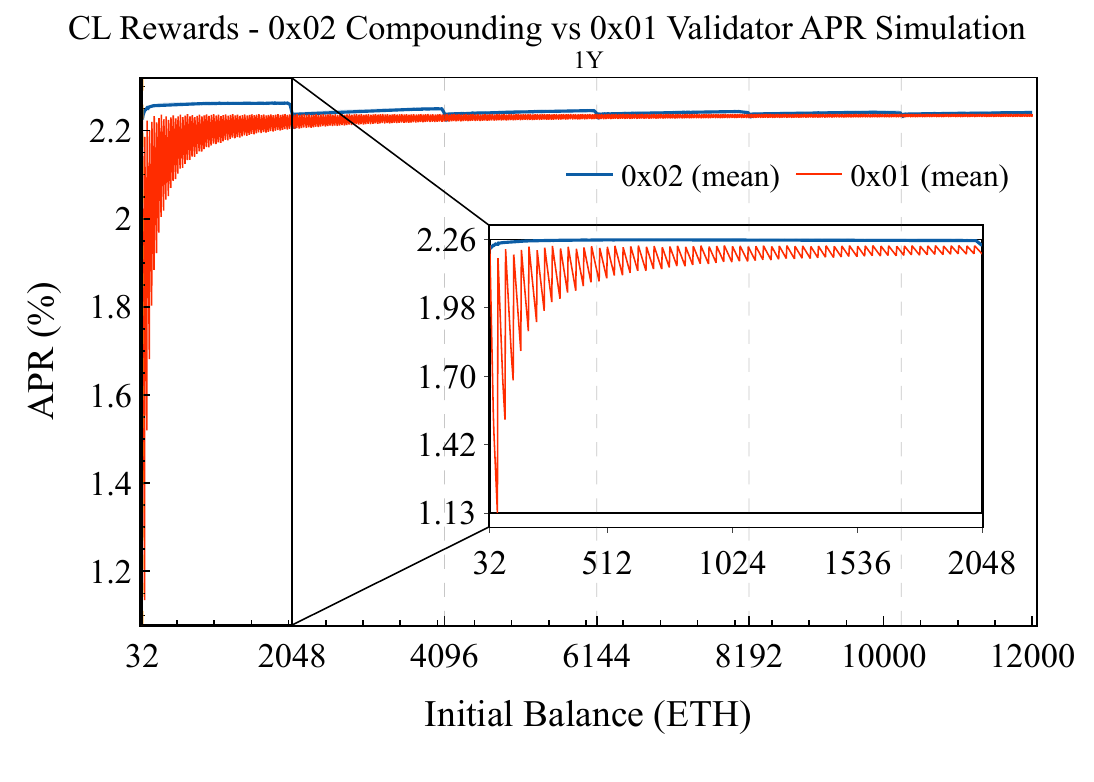}

    \caption{APR vs initial balance: \texttt{0x02} compounding validators maintain a consistent APR around 2.25\%, whilst \texttt{0x01} non-compounding validators show lower APR at smaller balances, gradually converging as balance increases.}
    \label{fig:apr-curve}
\end{figure}

To analyse how compounding scales with stake size, we simulate validator behaviour over a 1-year horizon for balances ranging from 32\,ETH to 10{,}240\,ETH using Monte Carlo simulation (200 runs per balance point, 96--99\% participation rate). Fig~\ref{fig:apr-curve} shows APR as a function of initial balance for both \texttt{0x02} and \texttt{0x01} setups, whilst Table~\ref{tab:apr-buckets} summarises the results by balance interval.

The compounding advantage is most pronounced for smaller balances typical of solo stakers, where \texttt{0x01} suffers from idle ETH sitting below the 32\,ETH activation threshold. As stake size increases, APR converges between both setups: large staking providers can allocate capital more efficiently across multiple validators, reducing the impact of remainder ETH.\footnote{In practice, validator entry and exit queues also affect performance (Fig~\ref{fig:validator-queue}); new validators may wait days or weeks to activate, making \texttt{0x02} compounding additionally attractive by avoiding queue delays.}

Based on these simulations, the CL APR uplift from compounding should range from +4.7\% for solo stakers (32--2{,}048\,ETH) to +0.3\% for large providers (8{,}192--10{,}240\,ETH). This suggests that \texttt{0x02} credentials should offer the greatest benefit to solo stakers and smaller operators, whilst large staking providers would gain only marginal improvement relative to their existing infrastructure.\footnote{Regarding Execution-layer rewards (priority fees and MEV): They are earned when selected as block proposer. Per EIP-7251~\cite{eip7251}, the probability of validator $v$ being selected is proportional to its effective balance:$\Pr_{v}(\text{proposer}) = \frac{B_{\text{eff},v}}{B_{\text{total}}}$ For equivalent total stake, aggregate proposer probability is identical regardless of credential type: a single \texttt{0x02} validator with 2{,}048\,ETH has the same expected proposals as 64 \texttt{0x01} validators at 32\,ETH each. However, as \texttt{0x02} effective balance grows through compounding, proposer probability increases proportionally, whilst \texttt{0x01} setups require activating new validators subject to entry queue delays (Fig~\ref{fig:validator-queue}).}

\begin{table}[b]
    \centering
    \caption{Monte Carlo simulation results (200 runs per balance point, 1-year horizon): Mean APR (\%) $\pm$ std and ETH rewards by balance bucket.}
    \label{tab:apr-buckets}
    \begin{tabular}{crrrrc}
        \toprule
        \textbf{}                  & \multicolumn{2}{c}{\textbf{$R_{\text{CL}}$ (ETH)}} & \multicolumn{2}{c}{\textbf{$\text{APR}_{\text{CL}}$ (\%)}} & \textbf{Uplift}                                                 \\
        \cmidrule(lr){2-3} \cmidrule(lr){4-5}
        \textbf{Balance (ETH)}     & \textbf{\texttt{0x01}}                             & \textbf{\texttt{0x02}}                                     & \textbf{\texttt{0x01}} & \textbf{\texttt{0x02}} & \textbf{(\%)} \\
        \midrule
        \textbf{32--2{,}048}       & 22.7                                               & 23.8                                                       & 2.17$\pm$.12           & 2.26$\pm$.01           & +4.7          \\
        \textbf{2{,}048--4{,}096}  & 69.2                                               & 69.8                                                       & 2.23$\pm$.01           & 2.25$\pm$.00           & +0.9          \\
        \textbf{4{,}096--6{,}144}  & 115.0                                              & 115.6                                                      & 2.23$\pm$.00           & 2.24$\pm$.00           & +0.5          \\
        \textbf{6{,}144--8{,}192}  & 160.8                                              & 161.4                                                      & 2.23$\pm$.00           & 2.24$\pm$.00           & +0.4          \\
        \textbf{8{,}192--10{,}240} & 203.8                                              & 204.4                                                      & 2.23$\pm$.00           & 2.24$\pm$.00           & +0.3          \\
        \bottomrule
    \end{tabular}
\end{table}

\subsection{Practical Implications}

Our simulations support the expected benefits: \texttt{0x02} compounding democratises returns for solo stakers (+4.7\% CL APR uplift for 32--2{,}048\,ETH) whilst offering marginal gains for large providers (+0.3\% for 8{,}192--10{,}240\,ETH). These assume idealised conditions: fixed network stake, high participation rates, no queue delays, and zero gas fees.

\textbf{Long-term holders} benefit most from \texttt{0x02}. Rewards compound automatically without gas fees, and the advantage grows over time.

\textbf{Staking-as-a-Service operators} face a nuanced trade-off. With \texttt{0x01}, excess balance above 32\,ETH is automatically swept at no cost. With \texttt{0x02}, accessing compounded rewards requires explicit partial withdrawal requests, incurring gas fees and queue delays. Operational costs may erode the theoretical APR advantage for providers needing frequent distributions.

\begin{figure}[!t]
    \centering
    \includegraphics[width=\columnwidth]{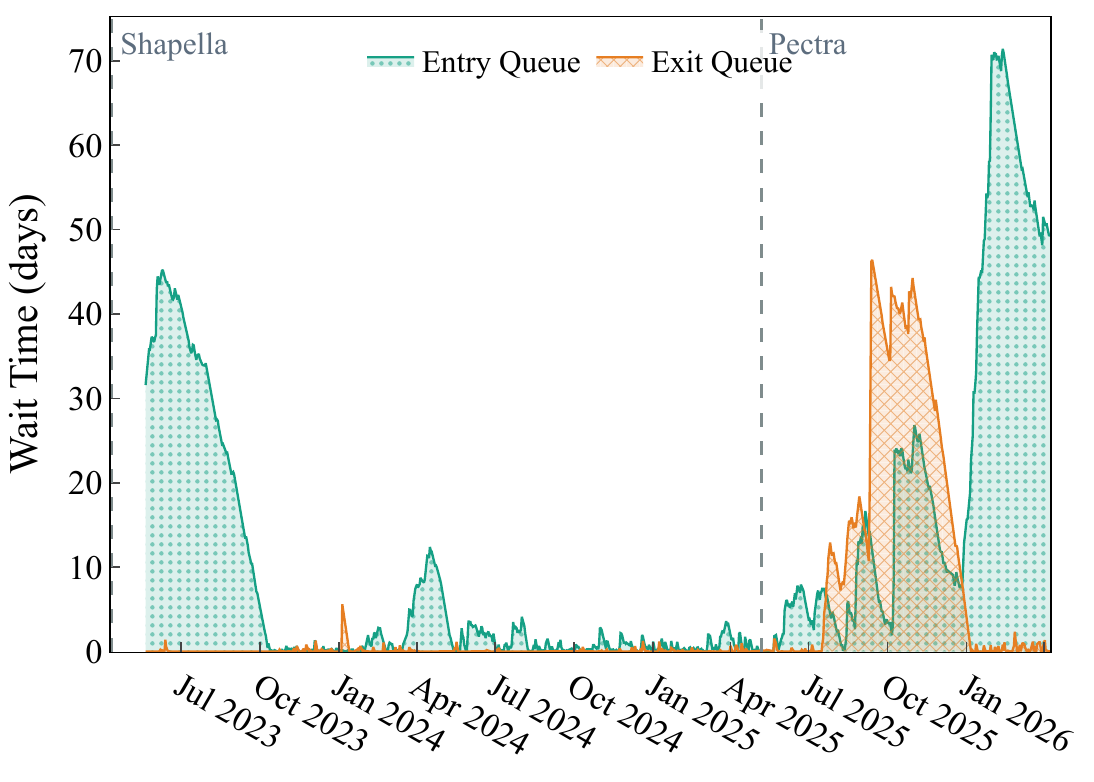}

    \caption{Validator entry and exit queue wait times since Shapella (Apr 2023). Peak entry wait reached 45 days (June 2023); exit queue peaked at 46 days (September 2025). Data: validatorqueue.com~\cite{validatorqueue}.}
    \label{fig:validator-queue}
\end{figure}

Queue dynamics further complicate this trade-off (Fig~\ref{fig:validator-queue}). For \texttt{0x01}, entry queues delay activation of new validators funded by accumulated rewards; exit queues exceeded 45 days during September 2025 when Kiln withdrew 1.6M\,ETH~\cite{kiln_exit_queue_2025}. For \texttt{0x02}, partial withdrawal requests face similar constraints.

In summary, the simulated APR uplift represents a lower bound for long-term holders. For operators requiring recurring distributions, the advantage may be smaller after accounting for gas costs and withdrawal delays. The following section examines these findings using on-chain data.

\section{Post-Pectra Empirical Analysis}
\label{sec:empirical}

Having established the theoretical framework and simulation results, we now turn to on-chain data from the Ethereum beacon chain to examine the realised performance and adoption patterns of \texttt{0x02} validators since the Pectra upgrade went live.

\begin{figure}[!t]
    \centering
    \includegraphics[width=\columnwidth]{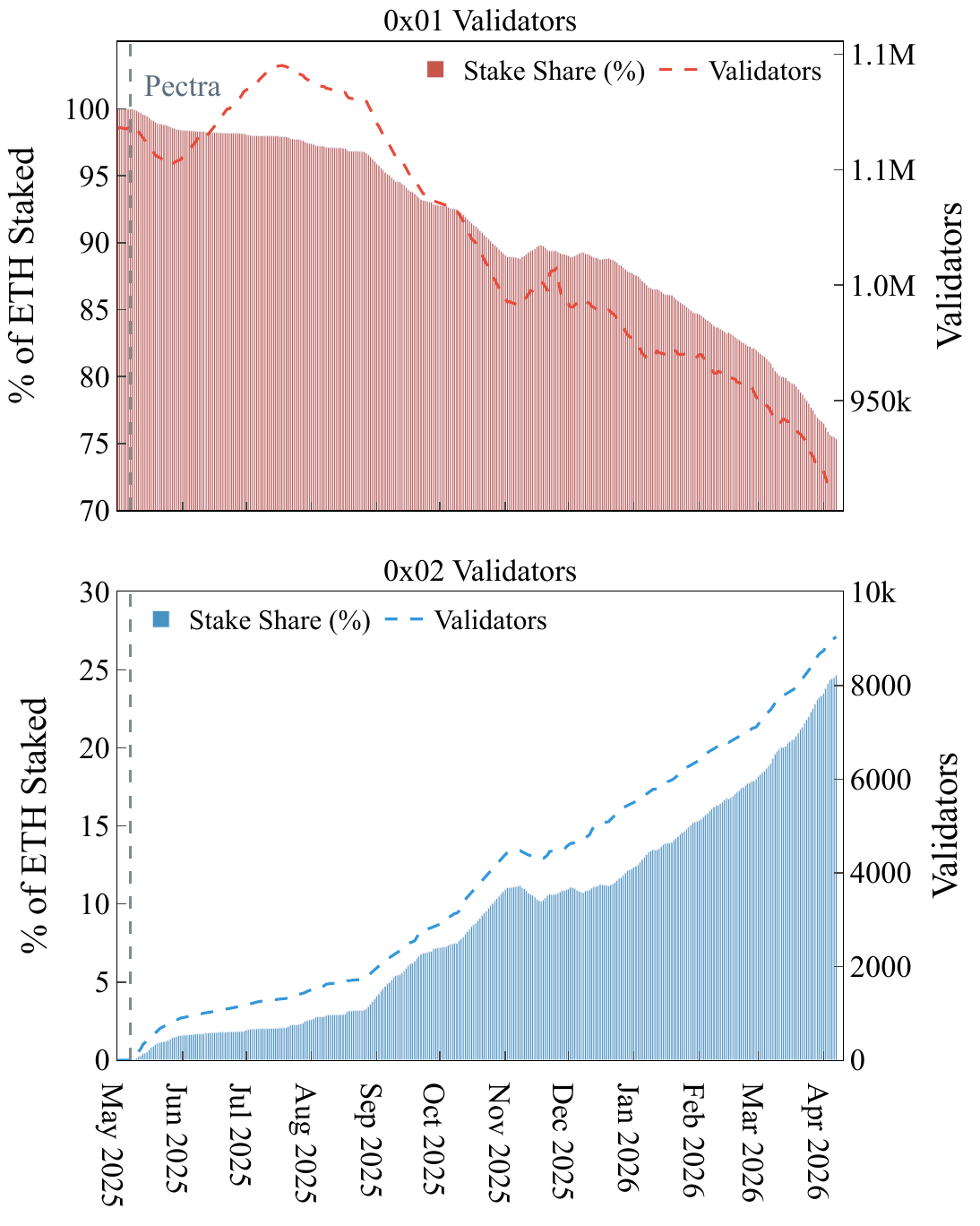}

    \caption{Validator adoption since Pectra launch: \texttt{0x01} validators (top) showing approximately 911{,}000 validators with 75\% stake share, and \texttt{0x02} validators (bottom) showing growth to 11{,}573 validators with 25\% stake share as of April 2026. Data source: Dune Analytics~\cite{dune_eth_staking}.}
    \label{fig:0x02-adoption}
\end{figure}

\begin{figure}[!t]
    \centering
    \includegraphics[width=\columnwidth]{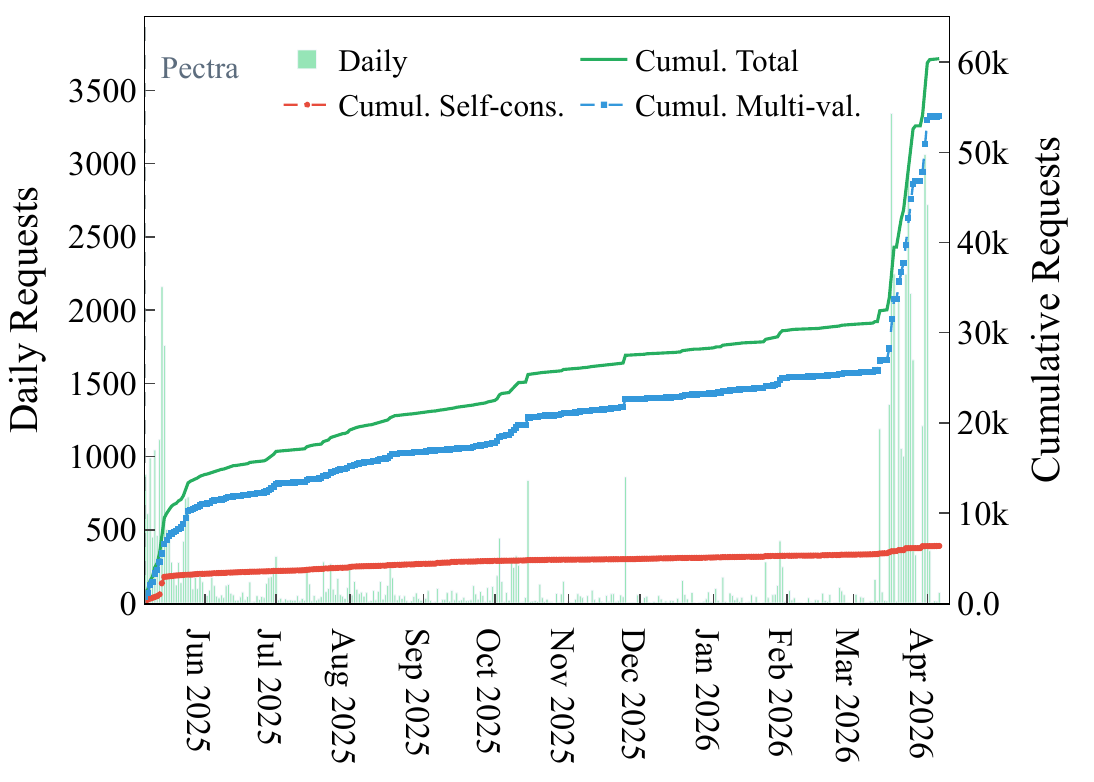}

    \caption{Consolidation requests since Pectra: daily volume (bars), cumulative total (solid line), and cumulative breakdown by type, multi-validator (blue) and self-consolidation (red). Data source: Dune Analytics~\cite{dune_pectra_consolidations}.}
    \label{fig:consolidation}
\end{figure}

\begin{figure}[!t]
    \centering
    \includegraphics[width=\columnwidth]{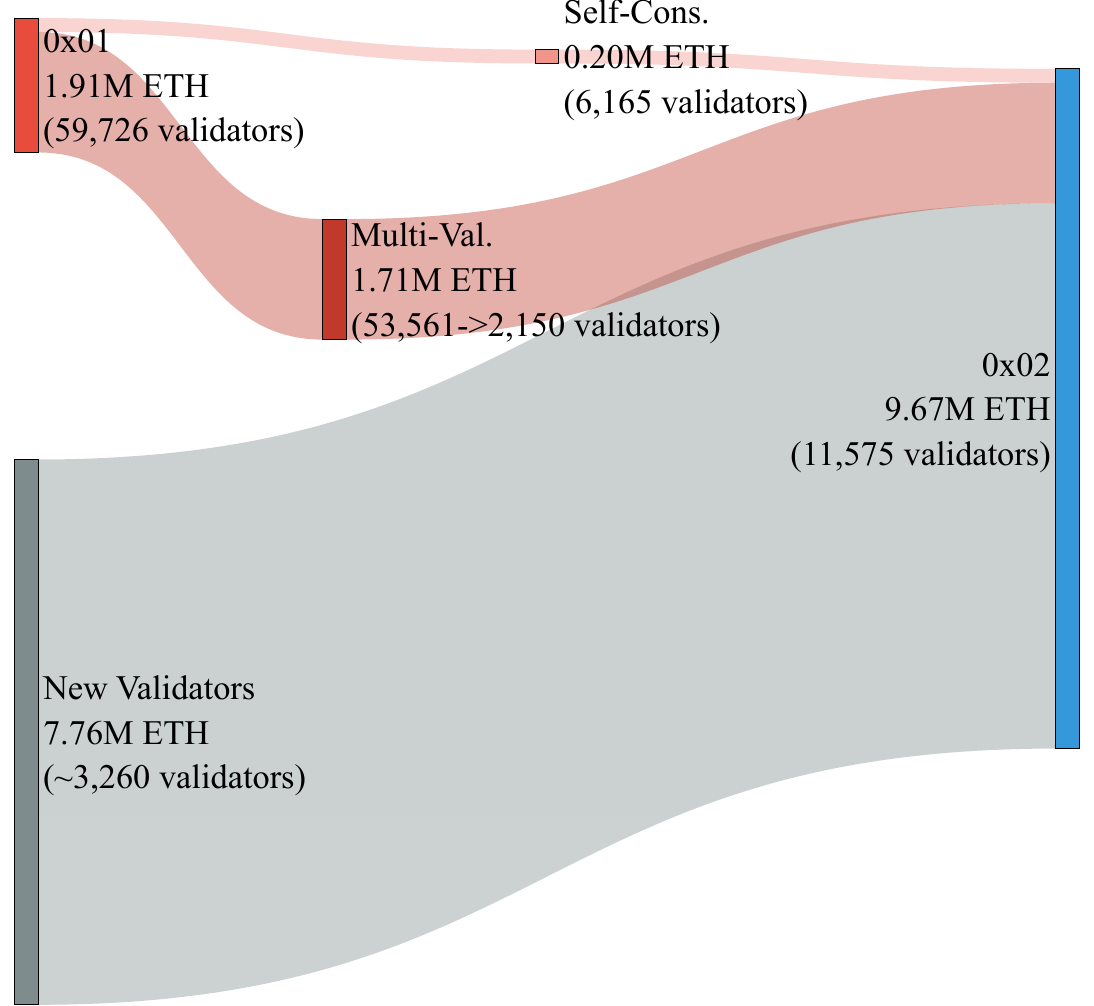}
    \vspace{-2mm}
    \caption{ETH flow to \texttt{0x02} validators (Apr 7, 2026). Consolidation contributes 1.91M\,ETH (197k ETH from 6{,}165 self-consolidations, 1.71M ETH from 53{,}561 multi-validator merges into 2{,}150 targets), whilst new direct deposits contribute 7.76M\,ETH across 3{,}260 validators. Total \texttt{0x02} stake: 9.67M ETH across 11{,}575 validators. Data: Dune Analytics~\cite{dune_pectra_consolidations}, beaconcha.in~\cite{beaconchain}.}
    \label{fig:consolidation-sankey}
\end{figure}

\subsection{\texttt{0x02} Validator Adoption}

As of the data collection period (April 7, 2026), the beacon chain hosts approximately 38.8\,million staked ETH across 922{,}721 active validators~\cite{dune_eth_staking,beaconchain}. This total comprises 9{,}496 legacy \texttt{0x00} validators (1.0\%), 901{,}650 \texttt{0x01} validators (97.7\%), and 11{,}575 \texttt{0x02} validators (1.3\%). Following our convention (Section~\ref{sec:withdrawal-credentials}), we exclude \texttt{0x00} credentials from the analysis, yielding a \texttt{0x01}/\texttt{0x02} subset of 913{,}225 validators used in all subsequent figures and statistics.\footnote{Small numerical differences across figures (e.g., 911{,}146 vs 901{,}650 for \texttt{0x01}) reflect whether \texttt{0x00} validators are grouped with \texttt{0x01} (as non-compounding) or excluded entirely. The total staked ETH is unaffected.} Fig~\ref{fig:staked_eth_intro} illustrates the broader staking landscape since Shapella, showing that 9.7M\,ETH (25\% of total stake) has migrated to \texttt{0x02} credentials in the eleven months following Pectra.

Fig~\ref{fig:0x02-adoption} provides a detailed view of this adoption trajectory. The top panel tracks total ETH staked in \texttt{0x02} validators, whilst the bottom panel shows the number of \texttt{0x02} validators and their share of total staked effective balance. Adoption grew steadily, reaching 11{,}575 validators and approximately 25\% of total effective balance by April 2026.

This adoption pace, whilst accelerating, remains incomplete relative to the network's scale. With 75\% of staked ETH still on \texttt{0x01} credentials eleven months post-Pectra, the transition continues gradually. The following subsections examine who is adopting and through which mechanisms.

\subsection{Validator Consolidation}
\label{sec:consolidation}

EIP-7251 enables operators to merge multiple validators into a single \texttt{0x02} credential with effective balance up to 2{,}048\,ETH~\cite{eip7251}. Unlike new deposits, consolidation represents existing stake reorganisation, providing insight into how operators restructure fleets post-Pectra. Consolidation requests are 96-byte transactions sent to a predeploy contract at \href{https://etherscan.io/address/0x0000BBdDc7CE488642fb579F8B00f3a590007251}{\texttt{0x00..7251}}, processed via a dedicated queue with source validators exiting through standard churn limits~\cite{eth_consensus_specs_electra}.

Fig~\ref{fig:consolidation} presents consolidation activity since Pectra launch. Daily volume averaged 186 requests per day across the observation period, but the cumulative curve reveals a clearly non-uniform pattern: activity remained low through the first ten months, then accelerated sharply in March 2026. Daily requests peaked at 3{,}343 in mid-March, and the three busiest days (3{,}343, 3{,}062 and 2{,}971 requests) all fell within that same month. Roughly half of the total 60{,}405 consolidation requests occurred in this final one-month window, more than doubling the 28{,}195 cumulative total observed through December 2025. Two modes emerge across the full period: \emph{multi-validator consolidation} (merging multiple validators) dominates at 89.5\% (54{,}049 requests), whilst \emph{self-consolidation} (upgrading credentials without merging) accounts for 10.5\% (6{,}356 requests).

Fig~\ref{fig:consolidation-sankey} illustrates ETH flow to \texttt{0x02} validators. Consolidation contributes 1.91M\,ETH (20\% of total \texttt{0x02} stake), whilst new deposits contribute 7.76M\,ETH (80\%). With consolidation accounting for only 20\% of the 9.67M\,ETH on \texttt{0x02}, new deposits continue to dominate adoption.

This suggests most \texttt{0x02} adoption comes from fresh capital rather than existing validators migrating. The dominance of multi-validator consolidation (89.5\% of requests, 1.71M\,ETH) indicates that operators who migrate prioritise reducing management overhead by aggregating stake, whilst self-consolidation (10.5\%, 197k\,ETH) captures the smaller share of pure credential upgrades.

\begin{figure}[!t]
    \centering
    \includegraphics[width=\columnwidth]{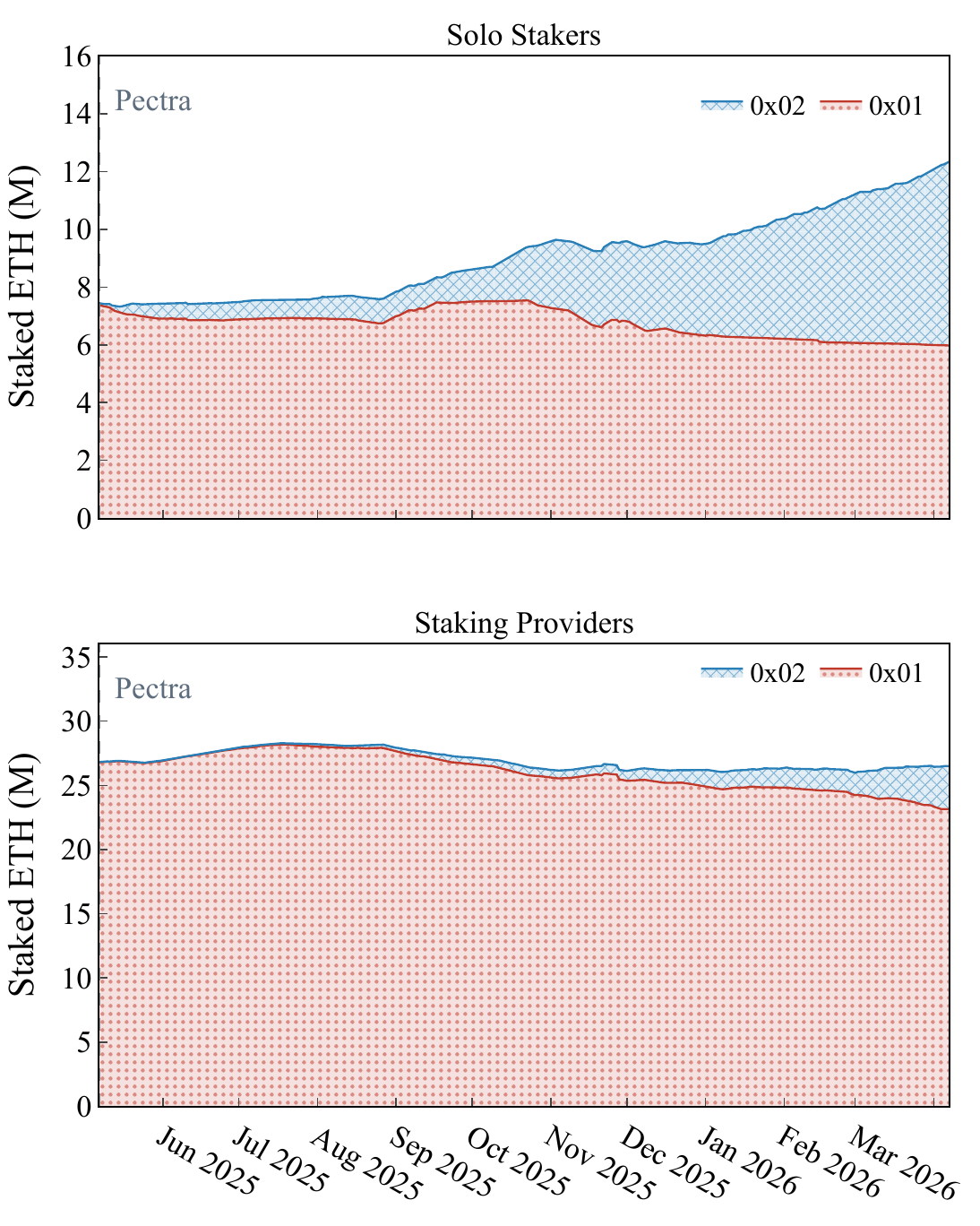}

    \caption{Staked ETH distribution by credential type for solo stakers (top) and staking providers (bottom) since Pectra launch. Solo stakers show higher \texttt{0x02} adoption relative to their stake, while staking providers remain predominantly on \texttt{0x01}. Data source: Dune Analytics~\cite{dune_eth_staking}.}
    \label{fig:solo-vs-providers}
\end{figure}

\subsection{Solo Stakers vs Staking Providers Adoption}

\begin{figure}[!t]
    \centering
    \includegraphics[width=\columnwidth]{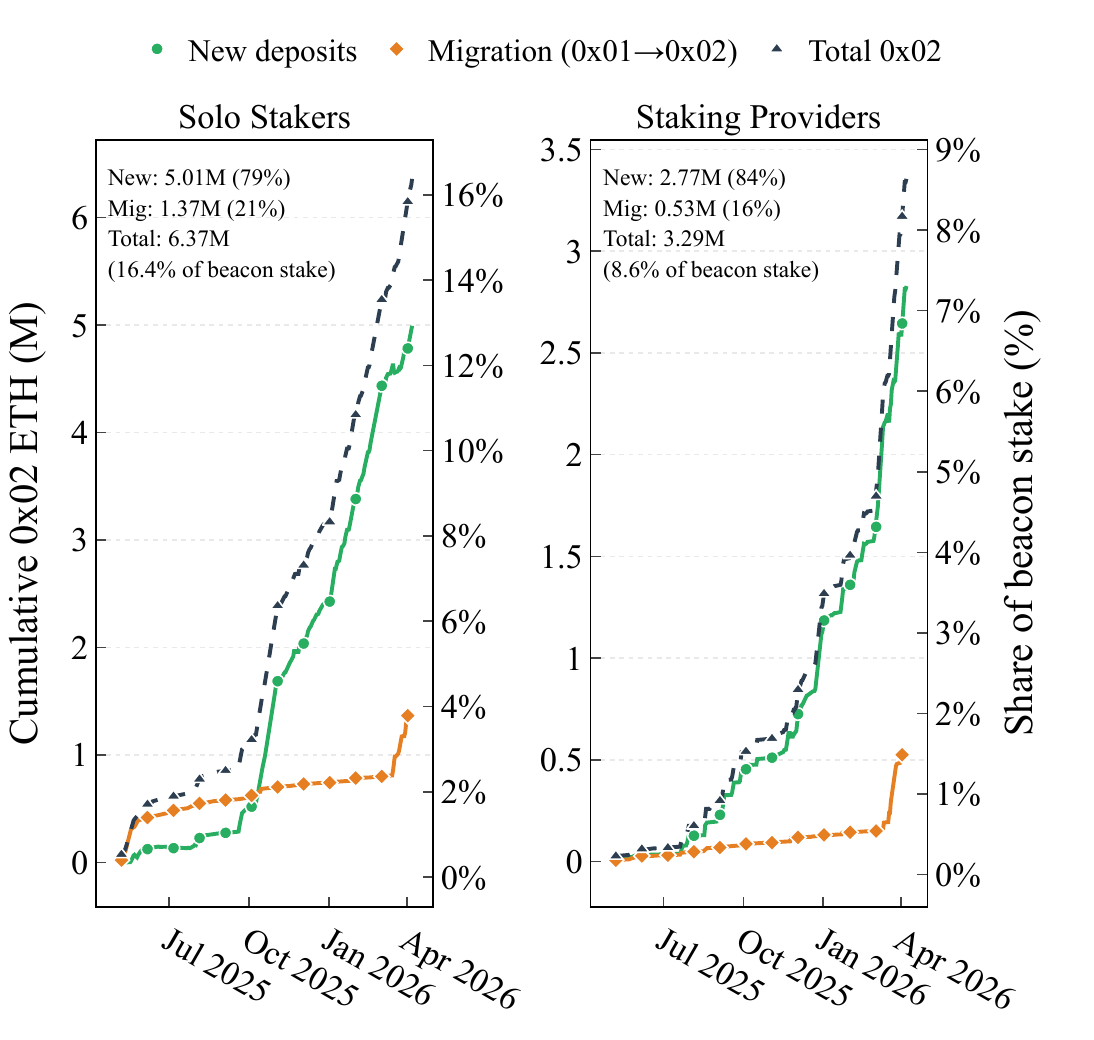}
    \caption{Cumulative \texttt{0x02} ETH per staker category, decomposed into new deposits (green) and \texttt{0x01}$\to$\texttt{0x02} migration (orange). Left y-axis: absolute ETH (M); right y-axis: same series as a share of total beacon-chain stake (\%). Solo and Providers \texttt{0x02} reach 16.4\% and 8.5\% of the 38.8M\,ETH beacon stake respectively (25\% total). Cut-off: 7~Apr~2026.}
    \label{fig:adoption-decomposed}
\end{figure}

The beacon-chain protocol does not record who operates each validator, so the ``Solo Stakers'' / ``Staking Providers'' split relies on an external attribution. We use the community-maintained Dune Analytics validator labels~\cite{dune_eth_staking}, which map individual validator indices to known staking entities (Lido, Coinbase, Binance, Kraken, Rocket Pool, EigenLayer operators, exchange platforms, LSTs, and so on). A validator that appears in any labelled entity set is counted as a \emph{Staking Provider}; any validator that does not appear in any labelled set is counted as a \emph{Solo Staker}. This is a conservative solo-staker definition: it catches genuine home stakers but also absorbs any professional operator whose validators have not yet been tagged by the community, so the Solo count is a slight upper bound. We apply the same mapping throughout the paper, including in Fig~\ref{fig:solo-vs-providers}, Fig~\ref{fig:adoption-decomposed}, and the Mann-Whitney cohorts of Section~\ref{sec:empirical-performance}.

We then split \texttt{0x02} adoption by staker category (Fig~\ref{fig:solo-vs-providers}, Fig~\ref{fig:adoption-decomposed}) and decompose each category's cumulative \texttt{0x02} stake into two components: \emph{new deposits} (validators entering the beacon chain post-Pectra and defaulting to \texttt{0x02}) and \emph{migration} (pre-existing \texttt{0x01} stake converted via consolidation). The migration component is anchored to the 1.91M\,ETH Sankey total of Fig~\ref{fig:consolidation-sankey} (59{,}726 unique source validators $\times$ 32\,ETH), so that the aggregate migrated ETH in Fig~\ref{fig:adoption-decomposed} exactly matches the real consolidation flow. Migration therefore accounts for only 1.91M\,ETH (19.8\%) of the 9.67M\,ETH cumulative \texttt{0x02} stake, with the remaining 7.76M\,ETH (80.2\%) coming from new direct deposits. New capital entering \texttt{0x02} dominates over conversion of existing \texttt{0x01} stake by a factor of four.

At the category level, solo stakers hold 6.37M\,ETH of \texttt{0x02} (16.4\% of the 38.8M\,ETH beacon stake) and staking providers hold 3.29M\,ETH (8.5\%), summing to the 25\% figure cited above. The two components follow very different trajectories in Fig~\ref{fig:adoption-decomposed}. New deposits (green) grow steadily for both categories throughout the 333-day window, with a mild acceleration towards the cut-off but no sharp inflection. Migration (orange) instead plateaus for most of the observation window (the cumulative line is nearly flat from mid-2025 through February 2026) before a late uptick in March~2026 that is visible in both subplots and coincides with the consolidation-event burst already noted in Fig~\ref{fig:consolidation-sankey}.

\subsection{Effective Balance Distribution}
\label{sec:effective-balance}

\begin{figure}[!t]
    \centering
    \includegraphics[width=\columnwidth]{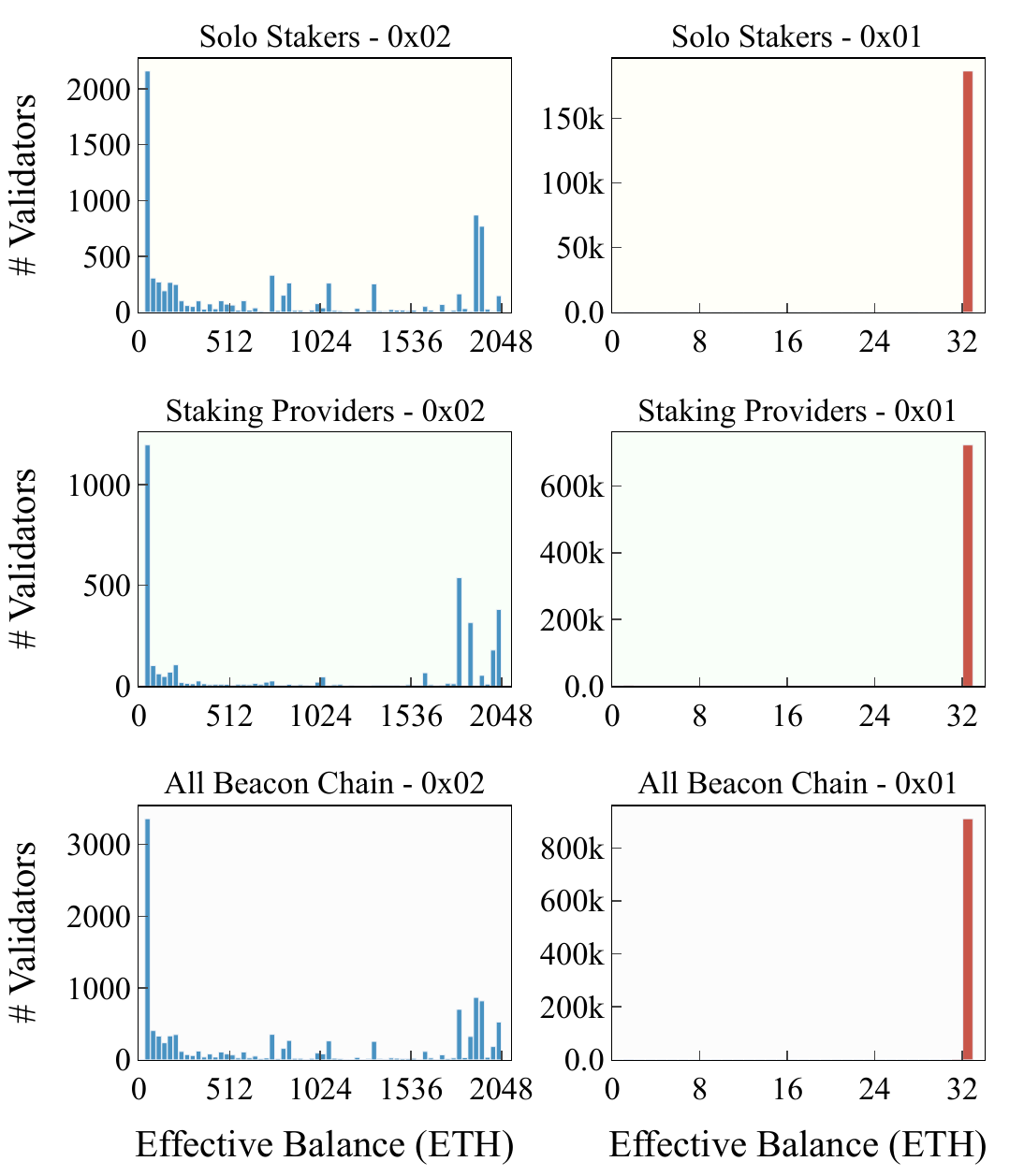}

    \caption{Effective balance distribution for \texttt{0x02} validators (left) and \texttt{0x01} validators (right), shown for all beacon chain validators (top), solo stakers (middle), and staking providers (bottom). Cut-off: 7~Apr~2026. Data sources: beaconcha.in~\cite{beaconchain} and Dune Analytics~\cite{dune_eth_staking}.}
    \label{fig:effective-balance}
\end{figure}

Beyond adoption rates, understanding how stake concentrates within \texttt{0x02} validators reveals distinct consolidation behaviours. Fig~\ref{fig:effective-balance} presents effective balance distributions for \texttt{0x02} validators (left) and \texttt{0x01} validators (right), segmented by validator type.

The \texttt{0x02} distribution exhibits a bimodal pattern: approximately 30\% cluster at low balances (32--64\,ETH), representing recently migrated validators, whilst 40\% concentrate above 1{,}024\,ETH, reflecting aggressive consolidation. Solo stakers show a broader spread (mean 842\,ETH), consistent with organic accumulation over time. Staking providers display a similar mean balance (815\,ETH), with validators concentrated either at entry-level or near the 2{,}048\,ETH cap.

This bimodal pattern suggests two distinct migration strategies: some operators upgrade credentials without consolidating, whilst others aggressively merge validators to reduce operational overhead.

\subsection{Empirical Validator Performance}
\label{sec:empirical-performance}

\begin{figure}[!t]
    \centering
    \includegraphics[width=\columnwidth]{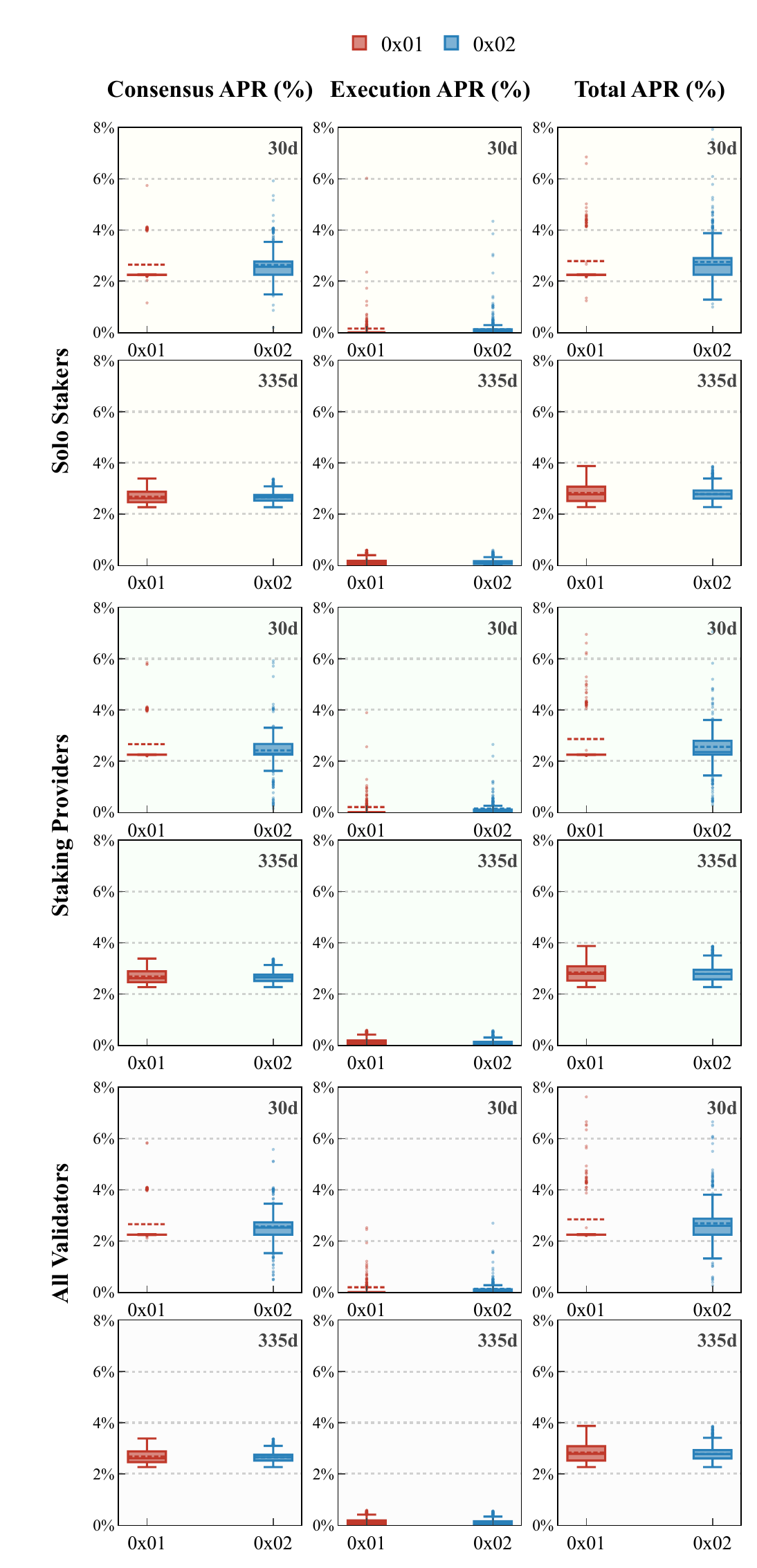}
    \caption{CL, EL, and Total APR distributions for \texttt{0x01} (red) and \texttt{0x02} (blue) validators, broken down by validator type. ``30d'' uses the 31-day reward accumulator (raw); ``335d'' (the full post-Pectra era, 7~May~2025 $\to$ 7~Apr~2026) uses the 365-day accumulator time-adjusted to each validator's observed history ($\geq$7 days) and 5/95 winsorised within each group, matching Table~\ref{tab:apr-windows} and the Mann-Whitney test cohort. Cut-off: 7~Apr~2026~\cite{beaconchain}.}
    \label{fig:rewards-apr-combined}
\end{figure}

We analyse APR data from \textbf{913{,}225 active validators} (901{,}650 \texttt{0x01} and 11{,}575 \texttt{0x02}) ending 7 April 2026~\cite{beaconchain}, matching the \texttt{0x01}/\texttt{0x02} subset introduced in Section~\ref{sec:empirical}. beaconcha.in exposes per-validator reward accumulators for four fixed rolling windows (1, 7, 31 and 365~days). To assess whether the compounding advantage is sustained over longer horizons (R2.5), we use the two that map cleanly onto these accumulators: a \textbf{30-day window} (from the 31-day accumulator) and a \textbf{335-day window} corresponding to the exact post-Pectra era (7~May~2025 $\to$ 7~Apr~2026). For the 335-day window we use a time-adjusted annualisation of the 365-day accumulator, $\text{APR} = (\text{cl}_{365d} / B_{\text{eff}}) \times 365 / \min(d, 365)$, where $d$ is the validator's days of active history; we require $d \geq 7$ to avoid extreme extrapolation. Fig~\ref{fig:rewards-apr-combined} presents the APR distributions for both windows and Table~\ref{tab:apr-windows} summarises medians.

CL APR here sums the attestation baseline ($\approx 2.24\%$/yr at 38.8M\,ETH staked) with proposer and sync-committee rewards, so the per-validator distribution has a long right tail driven by proposer luck and sync-committee selection; an ``average'' \texttt{0x01} validator earns approximately $2.61\%$ once annualised over 365 days.\footnote{Full component breakdown: (i) \emph{attestation rewards} each epoch for voting on source, target and head, providing the $\approx 2.24\%$ baseline; (ii) \emph{proposer rewards}, equal to $1/8$ of the attestation rewards in the block, earned roughly $\sim 2.85$ times per year per validator at 921k active validators; (iii) \emph{sync-committee rewards}, earned during a $\sim$27-hour slot every 256 epochs by 512 randomly-chosen validators (roughly once every six years per validator in expectation).}

\begin{figure}[!t]
    \centering
    \includegraphics[width=\columnwidth]{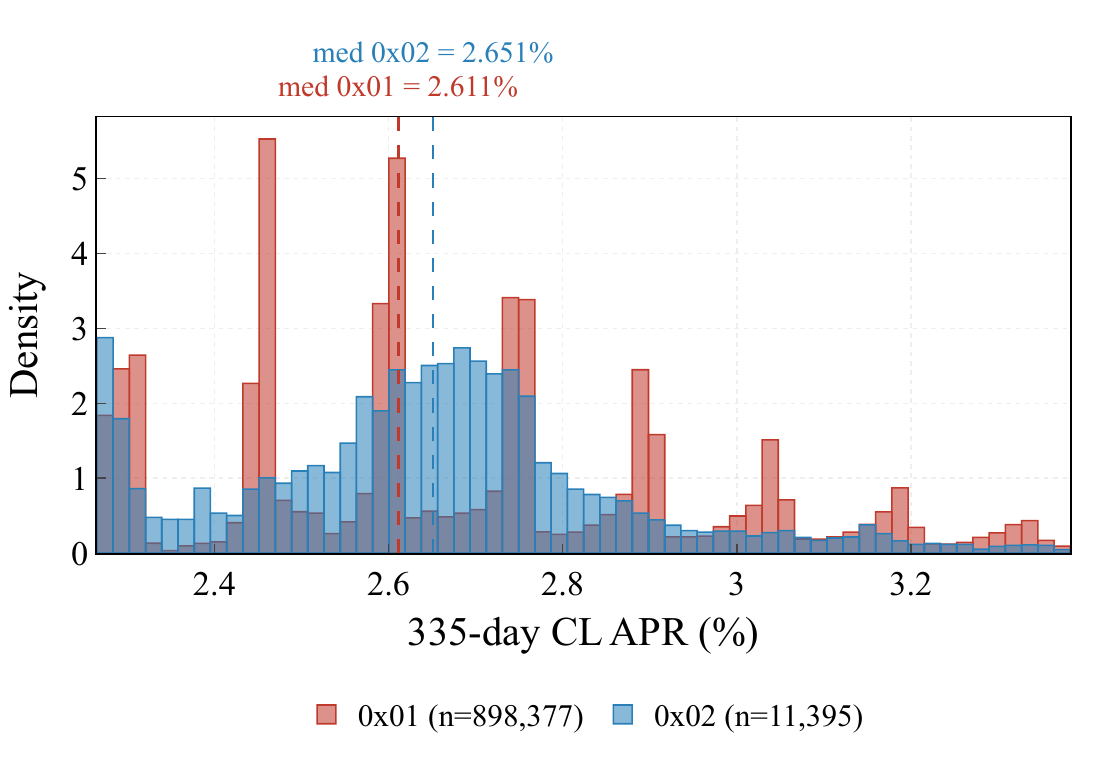}
    \caption{Overlaid histograms of 335-day CL APR for \texttt{0x01} (red) and \texttt{0x02} (blue), both winsorised at the 5/95 percentiles; dashed vertical lines mark the per-group medians. The \texttt{0x02} distribution is slightly right-shifted of the \texttt{0x01} baseline. Both distributions are markedly non-Gaussian and carry heavy right tails driven by proposer-reward and sync-committee excursions (the raw \texttt{0x01} distribution has excess kurtosis of $+10.9$, an order of magnitude above a Normal).}
    \label{fig:apr-335-hist}
\end{figure}

\begin{table}[!t]
    \centering
    \caption{CL APR medians by credential type. $\Delta$ is the relative median difference. The 30-day row uses the \texttt{cl\_31d} accumulator (raw, covers 89\% of \texttt{0x02} stake); the 335-day row covers the full post-Pectra era (7~May~2025 $\to$ 7~Apr~2026) via the time-adjusted 365-day accumulator, 5/95 winsorised to remove lower-tail annualisation noise from recent-deposit validators (same cohort as the Mann-Whitney test in Table~\ref{tab:apr-335-test}).}
    \label{tab:apr-windows}
    \setlength{\tabcolsep}{3pt}
    \begin{tabular}{@{}lccrrr@{}}
        \toprule
        & \multicolumn{2}{c}{\textbf{Median $\pm$ Std (\%)}} & & & \\
        \textbf{Window} & \textbf{\texttt{0x01}} & \textbf{\texttt{0x02}} & \textbf{$\Delta$} & $n_{\texttt{0x02}}$ & \textbf{0x02 stake} \\
        \midrule
        30\,d   & \AprThirtyMedOne\ $\pm$ \AprThirtyStdOne & \AprThirtyMedTwo\ $\pm$ \AprThirtyStdTwo & \AprThirtyDelta & \NThirtyTwo & \EthThirtyTwo \\
        335\,d  & \AprWindowMedOne\ $\pm$ \AprWindowStdOne & \AprWindowMedTwo\ $\pm$ \AprWindowStdTwo & \AprWindowDelta & \NWindowTwo & \EthWindowTwo \\
        \bottomrule
    \end{tabular}
\end{table}

Our simulations predicted a $+4.7\%$ relative CL APR uplift for small balances (32--2{,}048\,ETH), decreasing to $+0.3\%$ for large operators (Table~\ref{tab:apr-buckets}). We now compare these predictions against the empirical record, first descriptively at the two time horizons and then via a formal hypothesis test on the long-horizon window.

\paragraph{30-day window.} Over the most recent 30-day window, \texttt{0x02} validators show an aggregate CL APR advantage of \textbf{$+$12.3\%} relative to \texttt{0x01} (median \textbf{2.525\%} vs \textbf{2.249\%}), but this advantage is highly uneven across staker types: solo stakers show a \textbf{$+$13.9\%} uplift (median \textbf{2.561\%} vs \textbf{2.248\%}), whilst staking providers are essentially at parity with the attestation baseline (\textbf{$+$1.1\%}; \textbf{2.274\%} vs \textbf{2.249\%}). The 30-day cohort covers 10{,}763 of the 11{,}575 active \texttt{0x02} validators and represents 8.65M\,ETH out of the 9.67M\,ETH total \texttt{0x02} stake, so these statistics reflect essentially the full \texttt{0x02} population. The \texttt{0x02} distribution is notably wider than \texttt{0x01} (std 0.92\% vs 1.09\%) because proposer-reward luck over a 31-day window produces large per-validator variance. The observed solo uplift ($+$13.9\%) exceeds the simulated prediction ($+$4.7\% for 32--2{,}048\,ETH). The mechanism is proposer-reward variance, not compounding: solo \texttt{0x02} validators are disproportionately consolidated to higher effective balances ($B_{\text{eff}}$ median 480\,ETH in the 30-day cohort), so their proposer selection probability (which is proportional to effective balance) is roughly $15\times$ that of a 32\,ETH validator, and a lucky month of block proposals lands them on the right tail of an already heavy-tailed rewards distribution. \textbf{The direct proof of this interpretation is the convergence at longer horizons}: the $+$13.9\% solo lead collapses to $+$1.5\% at 335~days (Table~\ref{tab:apr-windows}). A genuine compounding gap would persist or grow with time; a proposer-variance gap averages out, which is exactly what the data show.

\paragraph{335-day (full post-Pectra era) window.} A naive ``last 365 days'' filter would only cover the $\approx$20\% of \texttt{0x02} stake that belongs to validators activated before Pectra, because Pectra is only $\approx$11 months old and fresh-deposit \texttt{0x02} validators do not yet have a full year of history. To avoid that bias, we use the longest window the data actually spans (335 days, the exact gap between Pectra activation and the snapshot cut-off) and annualise each validator's own observed history by scaling the 365-day accumulator by $365 / \min(d, 365)$ (with $d$ the days of active history and $d \geq 7$). For a validator active only a few days, $d$ is tiny and the scaling factor $365/d$ is correspondingly large (e.g., $d=10$ gives a factor of $36.5$): tiny fluctuations in the raw reward accumulator are amplified into large per-validator APR swings, and a disproportionate share of recent \texttt{0x02} deposits end up in the left tail of the empirical distribution. To keep these scaling artefacts from dominating the per-cohort medians, we apply a 5/95 winsorisation within each group before reporting any statistics (Table~\ref{tab:apr-windows}, 335d row). Once we average over each validator's full observed history and winsorise, \textbf{the 30-day advantage shrinks dramatically}: \texttt{0x01} median CL APR settles at \textbf{2.61\%} (vs.~\textbf{2.25\%} at 30 days) as proposer-reward luck averages out, and \texttt{0x02} settles at \textbf{2.65\%}, a small \textbf{$+1.5\%$} relative gap rather than the inflated \textbf{$+12.3\%$} 30-day lead. The 30-day figure was a \emph{short-window variance artefact}: high-$B_{\text{eff}}$ validators (disproportionately \texttt{0x02}) have a higher proposer-selection probability, and in a 31-day window the proposer-reward component dominates any compounding-related signal. Over longer horizons the proposer component averages toward its expectation for every validator, and the distributions converge onto a \textbf{small residual \texttt{0x02} advantage}.

We formalise the 335-day comparison with a rank-based hypothesis test on three cohorts $c \in \{\text{All},\text{Solo},\text{Providers}\}$: $H_0^{(c)}$ states that the 335-day CL APR distributions of \texttt{0x02} and \texttt{0x01} validators are identical, and $H_1^{(c)}$ (two-sided) that they differ, tested at $\alpha = 0.05$.

Picking the right test comes down to the shape of the two underlying samples. A Student $t$-test assumes approximately Normal samples and a common variance, and its Welch variant relaxes the common-variance assumption but still relies on near-Normal distributions. Neither holds here: the raw \texttt{0x01} distribution carries a long right tail from the lumpy proposer-reward and sync-committee components of CL APR, and even after 5/95 winsorisation it retains a positive skew of $\NormSkew$ (Fig~\ref{fig:apr-335-hist}). A Shapiro-Wilk normality test on the winsorised \texttt{0x01} sample rejects the Normal null decisively ($W = \NormW$, $p \approx \NormPval$). On top of that, the 335-day \texttt{0x02} sample has a much larger spread than \texttt{0x01} before winsorisation, as the time-adjusted annualisation factor amplifies reward-accumulator noise on recent deposits. We therefore use the two-sided \textbf{Mann-Whitney $U$} (Wilcoxon rank-sum), a nonparametric test that operates on the joint ranks of the two samples and tests the null hypothesis $\Pr(X_{\texttt{0x02}} > X_{\texttt{0x01}}) = \Pr(X_{\texttt{0x01}} > X_{\texttt{0x02}})$ without any distributional assumption. With $n_{\texttt{0x01}}$ in the $10^5$--$10^6$ range the test has enormous statistical power, which is exactly the regime in which one needs to separate the two questions a hypothesis test mixes together: \emph{is there a real difference} and \emph{how large is it}. We rely on the $p$-value for the first question, as a binary reject-or-accept flag on the null, and on the per-cohort median difference $\Delta_{\text{med}} = \mathrm{med}_{\texttt{0x02}} - \mathrm{med}_{\texttt{0x01}}$ (in percentage points) as the effect-size statistic that quantifies the magnitude of the shift, since with samples this large the $p$-value alone saturates for any non-negligible effect.

\begin{table}[!t]
    \centering
    \caption{Mann-Whitney $U$ test of 335-day CL APR, \texttt{0x02} vs \texttt{0x01}, by cohort. $\Delta_{\text{med}}$ is the raw median difference (\texttt{0x02}$-$\texttt{0x01}) in percentage points; $p$ is the two-sided Mann-Whitney $U$ $p$-value. All three cohorts reject $H_0$ at $\alpha = 0.05$. Cut-off: 7~Apr~2026.}
    \label{tab:apr-335-test}
    \footnotesize
    \setlength{\tabcolsep}{2pt}
    \resizebox{\columnwidth}{!}{%
    \begin{tabular}{@{}lrrccrr@{}}
        \toprule
        \textbf{Cohort} & $n_{\texttt{01}}$ & $n_{\texttt{02}}$ & \textbf{Med\textsubscript{01}} (\%) & \textbf{Med\textsubscript{02}} (\%) & $\Delta_{\text{med}}$ (pp) & $p$ \\
        \midrule
        All Validators    & \NAllOne  & \NAllTwo  & \MedAllOne  & \MedAllTwo  & $\DeltaAll$  & $\PAll$  \\
        Solo Stakers      & \NSoloOne & \NSoloTwo & \MedSoloOne & \MedSoloTwo & $\DeltaSolo$ & $\PSolo$ \\
        Staking Providers & \NProvOne & \NProvTwo & \MedProvOne & \MedProvTwo & $\DeltaProv$ & $\PProv$ \\
        \bottomrule
    \end{tabular}%
    }
\end{table}

\begin{figure}[!t]
    \centering
    \includegraphics[width=\columnwidth]{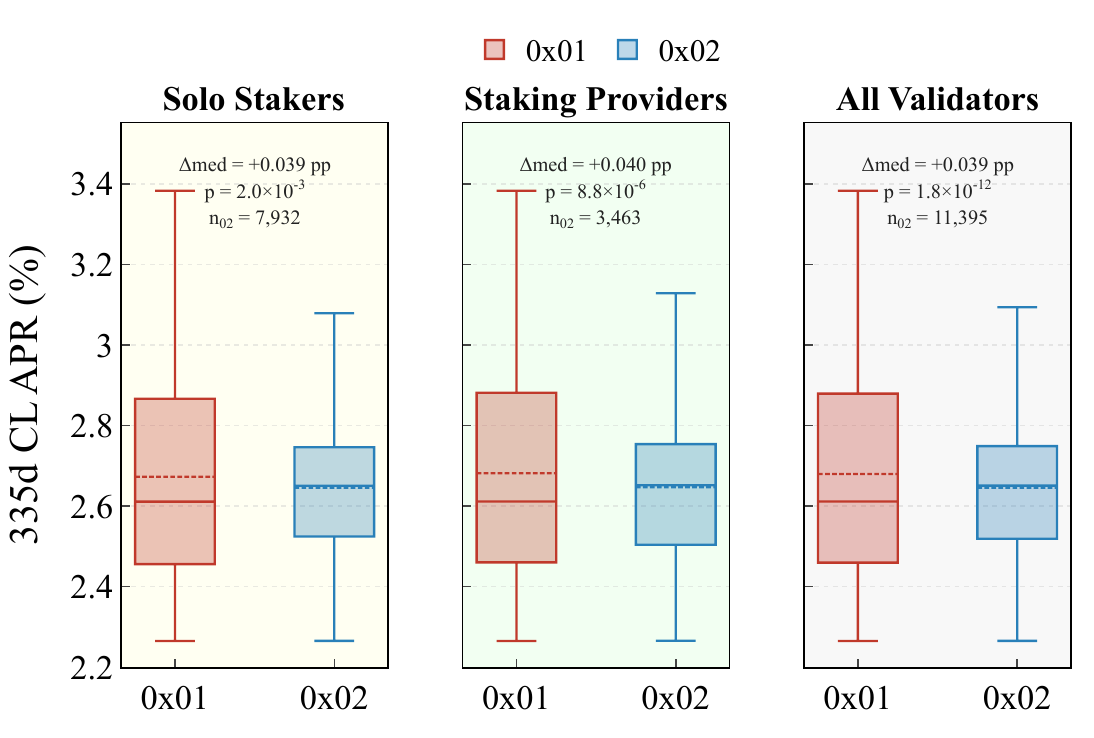}
    \caption{335-day CL APR distributions for \texttt{0x01} (red) and \texttt{0x02} (blue) by cohort. Box statistics are winsorised at the 5/95 percentiles; each panel reports the median difference $\Delta_{\text{med}} = \mathrm{med}_{\texttt{0x02}} - \mathrm{med}_{\texttt{0x01}}$, the Mann-Whitney $U$ $p$-value, and $n_{\texttt{02}}$.}
    \label{fig:apr-boxplot-335}
\end{figure}

Table~\ref{tab:apr-335-test} and Fig~\ref{fig:apr-boxplot-335} report the results. At $\alpha = 0.05$, \textbf{$H_0$ is rejected in all three cohorts}:
\needspace{5\baselineskip}%
\begin{itemize}
    \setlength\itemsep{0pt}
    \setlength\topsep{0pt}
    \setlength\partopsep{0pt}
    \setlength\parskip{0pt}
    \item \textbf{All Validators}: $\Delta_{\text{med}} = \DeltaAll$\,pp, $p = \PAll \ll 0.05$ \textbf{$\Rightarrow$ reject $H_0$}.
    \item \textbf{Solo Stakers}: $\Delta_{\text{med}} = \DeltaSolo$\,pp, $p = \PSolo < 0.05$ \textbf{$\Rightarrow$ reject $H_0$}.
    \item \textbf{Staking Providers}: $\Delta_{\text{med}} = \DeltaProv$\,pp, $p = \PProv \ll 0.05$ \textbf{$\Rightarrow$ reject $H_0$}.
\end{itemize}
The direction is \emph{positive in every cohort}: the 335-day CL APR of \texttt{0x02} is stochastically higher than \texttt{0x01} by \textbf{$\DeltaAllPP$\,pp} in absolute terms (about \textbf{$\DeltaAllRel$} relative), so $H_1$ is supported throughout.

This gap is about \textbf{one-third} of the \textbf{$+0.13$\,pp} uplift (\textbf{$+4.7\%$} relative) our simulation predicts for a 32--2{,}048\,ETH validator in Section~\ref{sec:scaling-effects}. The remaining two-thirds are consistent with a compounding mechanism operating on a \emph{multi-year horizon} that \textbf{11 months} of post-Pectra data have only begun to sample: at current $\sim$2.25\% yield, a 32\,ETH validator needs $\sim$\textbf{193 days} to trigger its first effective-balance increment, and most \texttt{0x02} validators have not yet crossed this threshold naturally. A reassessment after 2--3 years, or under the accelerated hysteresis proposed in EIP-8068~\cite{eip8068}, would better capture organic compounding.

Putting everything together, the empirical record \textbf{directionally confirms} the simulation's prediction that \texttt{0x02} compounding validators outperform \texttt{0x01} on CL APR: we observe a \textbf{small but statistically significant positive shift in every cohort we tested}, matching the sign and qualitative behaviour of Section~\ref{sec:scaling-effects}. Quantitatively, however, the empirical uplift is only about \textbf{a third} of the simulated multi-year steady-state: \textbf{$\DeltaAllPP$\,pp} (\textbf{$\DeltaAllRel$}) in the 335-day window against the simulation's \textbf{$+0.13$\,pp} (\textbf{$+4.7\%$}). This is a \textbf{partial match rather than a full one}, and the size of the gap is exactly what the hysteresis-dominated compounding mechanism predicts for an 11-month observation window. The simulation and the data are therefore \textbf{not in tension}; they describe the same effect at different points in its accrual trajectory, and the residual two-thirds are a prediction about the \emph{next two to three years} rather than a missing effect in our measurement.

\FloatBarrier

\section{Limitations}
\label{sec:threats}
Our findings are subject to several limitations. The simulation assumes a \textbf{fixed total network stake of 38.8M\,ETH} and a participation rate of 96--99\%, and \textbf{excludes proposer rewards}, which may affect the absolute APR levels.
The empirical analysis uses two observation windows: a 30-day window (from the \texttt{cl\_31d} reward accumulator) and a 335-day window corresponding to the full post-Pectra era (Pectra: 7~May~2025 $\to$ cut-off: 7~April~2026 $=$ 335 days), time-adjusted via the 365-day accumulator. The 335-day window is already the longest observation period the data permit, but \textbf{335 days is still a short horizon} for a compounding mechanism whose first effective-balance increment under current hysteresis rules takes ${\sim}$193 days at 2.25\% yield. A \textbf{reassessment after 2--3 full years post-Pectra} would be needed to observe the full simulated uplift directly; however, the mechanism itself is likely to change before then (Section~\ref{sec:upcoming-eips}), so any follow-up will need to re-baseline against the new rules.
Additionally, early adopters of \texttt{0x02} withdrawal credentials may represent more sophisticated operators, introducing potential \textbf{selection bias}; this concern is mitigated but not eliminated by the growing sample (11{,}575 \texttt{0x02} validators versus 901{,}650 \texttt{0x01}, total 913{,}225 in the analysis subset). Validator classification further relies on \textbf{heuristic tagging from the Dune validator tags dataset}, and misclassification, particularly of solo stakers, cannot be ruled out. Finally, \texttt{0x02} validators received a \textbf{higher number of block proposals during the 30-day measurement window}, inflating the short-horizon CL APR gap with proposer rewards that are independent of compounding; this is the reason we restrict the formal hypothesis test in Section~\ref{sec:empirical-performance} to the 335-day window.

\section{Conclusion}
\label{sec:conclusion}

This paper quantified the performance implications of \texttt{0x02} compounding validators introduced by EIP-7251. Simulations show \textbf{$+4.7\%$ relative CL APR uplift} for small balances (32--2{,}048\,ETH), diminishing to $+0.3\%$ for large providers. The formal 335-day hypothesis test over the full post-Pectra era confirms a \textbf{small but statistically significant \texttt{0x02} advantage} across all three cohorts (All, Solo, Providers), with median CL APR differences of \textbf{$+0.04$\,pp} ($\approx +1.5\%$ relative). The observed magnitude is about \textbf{one-third} of the simulated multi-year prediction, consistent with the hysteresis-dominated compounding mechanism being truncated by the \textbf{11-month observational window}.

Despite theoretical benefits, adoption has reached \textbf{25\% of staked ETH (9.7M\,ETH)} eleven months post-Pectra, with new deposits flowing into \texttt{0x02} and 0x01$\to$0x02 migration plateauing for most of the observation window before a \textbf{late uptick in March~2026}: both cumulative new \texttt{0x02} deposits and consolidation requests accelerate in the final month of our data, visible in Fig~\ref{fig:adoption-decomposed} and Fig~\ref{fig:consolidation}. \texttt{0x02} validators must explicitly request partial withdrawals (inducing gas fees) to access accumulated rewards, whilst \texttt{0x01} validators receive automatic sweeps of excess balance to their withdrawal address, using it as liquid cashflow. Staking providers designed their systems for fixed \texttt{32ETH} balances with automatic reward sweeps, and adapting to compounding yet locked rewards requires significant software changes. The small observed differences are not surprising: over short time horizons, compounding has little effect as balance hysteresis delays when rewards increase a validator's effective balance, only taking effect once the \textbf{33.25\,ETH threshold} is crossed. This reduces the immediate benefit of switching to \texttt{0x02}, making the change less urgent for most solo stakers. Faster balance updates, for example through adjustments to the hysteresis mechanism, could encourage earlier adoption. Even so, we observe some solo stakers switching to \texttt{0x02} early, effectively taking a long-term view on compounding rewards.

In contrast, \textbf{large institutional stakers are less inclined to switch to \texttt{0x02} credentials under the current rules, where compounding is effectively the only incentive on offer}. As shown in our simulations, the depth of their staking pools already lets them capture a compounding-like effect by activating new validators with earned rewards, which largely offsets the marginal benefit Pectra's in-place compounding would add. On top of that, \texttt{0x02} introduces operational drawbacks for large operators: rewards are no longer swept automatically, so partial withdrawals become a recurring software and cost item, and the locked-balance regime limits their ability to reallocate capital under the current hysteresis rules~\cite{eip8068}. Section~\ref{sec:upcoming-eips} discusses the draft EIPs (EIP-8068, EIP-8062, EIP-7804) that would address each of these frictions in turn and tilt the cost-benefit calculation towards migration for this cohort.

Ethereum's motivation for \texttt{0x02} is reducing validator count to improve network efficiency, as the legacy 32\,ETH cap creates congestion disproportionate to actual staked ETH. The approach favours organic migration by making \texttt{0x02} attractive rather than penalising \texttt{0x01}. However, \textbf{without significant APR uplift and consistent reward withdrawals, the economic case remains weak} for both solo stakers and providers. Ongoing discussions between core developers and staking providers~\cite{eth_acdc_consolidation_2025} continue to explore solutions balancing network efficiency goals with staker economics.

\paragraph{Upcoming Protocol Changes.}\label{sec:upcoming-eips}

Several draft EIPs under active discussion by Ethereum core developers would materially alter the \texttt{0x02} incentive structure analysed in this paper, and can be read as a direct response to the slow adoption pace documented here. \textbf{EIP-8068}~\cite{eip8068} addresses the hysteresis inefficiency at the core of our findings: it reduces the upward threshold from $+1.25$\,ETH to $+0.5$\,ETH and rounds the effective balance to the nearest integer, cutting the time to the first effective-balance increment from \textbf{${\sim}193$~days} (our simulation) to \textbf{${\sim}80$~days}. This would significantly \emph{strengthen the compounding advantage} and accelerate the timeline for observing it empirically. \textbf{EIP-8062} introduces a $0.05\%$ fee on automatic reward sweeps for \texttt{0x01} validators, adding a cost to remaining on legacy credentials that would directly shift the cost-benefit equation for migration. \textbf{EIP-7804} defines a new \texttt{0x03} request type enabling validators to update their withdrawal credentials without exiting and re-entering the validator set, \emph{removing a key friction point} for migration.

Together, these proposals signal a deliberate shift toward \textbf{making \texttt{0x02} more attractive} (EIP-8068), \textbf{making \texttt{0x01} less attractive} (EIP-8062), and \textbf{reducing migration friction} (EIP-7804). If adopted, these changes would likely accelerate \texttt{0x02} adoption beyond the current 25\% share. Additionally, \textbf{EIP-8071} addresses an exploit where the consolidation queue was used to bypass exit queues, which may explain some of the consolidation spikes in our data. Future research could reassess adoption dynamics and measure the realised compounding benefit after these protocol changes take effect.

\bibliographystyle{ACM-Reference-Format}
\bibliography{references}

\end{document}